\documentclass[aps,prd,twocolumn,superscriptaddress,groupedaddress]{revtex4}  % for review and submission
\usepackage{graphicx}  % needed for figures
\usepackage{dcolumn}   % needed for some tables
\usepackage{bm}        % for math
\usepackage{amssymb,amsmath,amsfonts}   % for math

\newcommand{\threep}{$3^{\,\prime}$\,}
\newcommand{\fivep}{$5^{\,\prime}$\,}

\newcommand{\beq}{\begin{equation}}
\newcommand{\eeq}{\end{equation}}
\newcommand{\bea}{\begin{eqnarray}}
\newcommand{\eea}{\end{eqnarray}}

\begin{document}

\title{Statistical inference of the generation probability of T-cell receptors from sequence repertoires}
\author{Anand Murugan}
\affiliation{Joseph Henry Laboratories, Princeton University, Princeton, New Jersey 08544 USA}
\author{Thierry Mora}
\affiliation{Laboratoire de physique statistique, UMR8550, CNRS and \'Ecole normale sup\'erieure, 24, rue Lhomond, 75005 Paris, France}
\author{Aleksandra M. Walczak}
\affiliation{Laboratoire de physique th\'eorique, UMR8549, CNRS and \'Ecole normale sup\'erieure, 24, rue Lhomond, 75005 Paris, France}
\author{Curtis G. Callan, Jr.}
\affiliation{Joseph Henry Laboratories, Princeton University, Princeton, New Jersey 08544 USA}
\affiliation{Simons Center for Systems Biology, Institue for Advanced Study, Princeton, New Jersey 08544 USA}

\begin{abstract}
Stochastic rearrangement of germline DNA by VDJ recombination is at the origin of immune system diversity. This process is implemented via a series of stochastic molecular events involving gene choices and random nucleotide insertions between, and deletions from, genes. We use large sequence repertoires of the variable CDR3 region of human CD4+ T-cell receptor beta chains to infer the statistical properties of these basic biochemical events.  Since any given CDR3 sequence can be produced in multiple ways, the probability distribution of hidden recombination events cannot be inferred directly from the observed sequences; we therefore develop a maximum likelihood inference method to achieve this end. To separate the properties of the molecular rearrangement mechanism from the effects of selection, we focus on non-productive CDR3 sequences in T-cell DNA. We infer the joint distribution of the various generative events that occur when a new T-cell receptor gene is created. We find a rich picture of correlation (and absence thereof), providing insight into the molecular mechanisms involved. The generative event statistics are consistent between individuals, suggesting a universal biochemical process. Our distribution predicts the generation probability of any specific CDR3 sequence by the primitive recombination process, allowing us to quantify the potential diversity of the T-cell repertoire and to understand why some sequences are shared between individuals. We argue that the use of formal statistical inference methods, of the kind presented in this paper, will be essential for quantitative understanding of the generation and evolution of diversity in the adaptive immune system.
\end{abstract}

\maketitle

% FIGURE 1

\newcommand{\panelone}{
\begin{figure*}
\noindent\includegraphics[width=\linewidth]{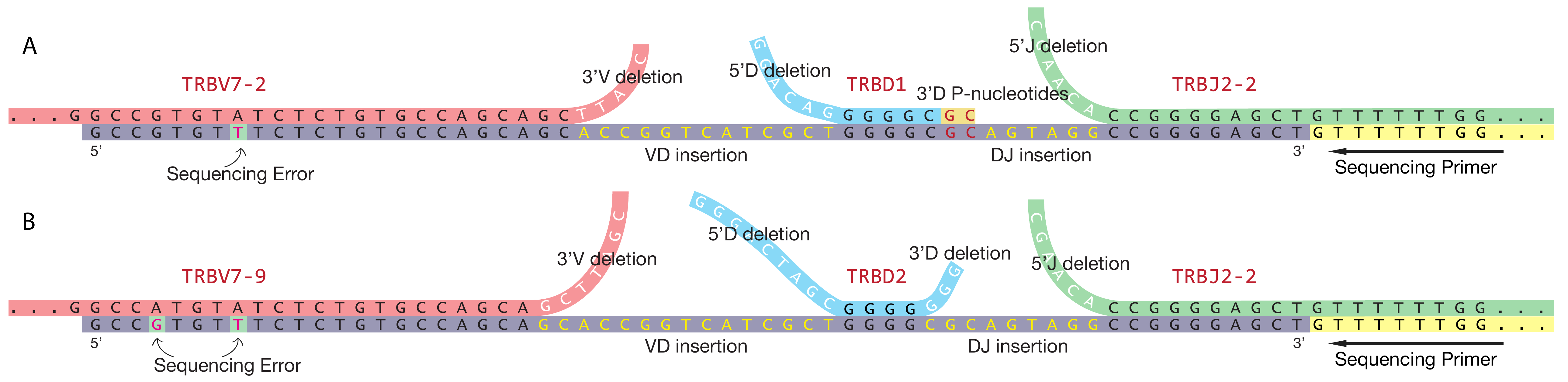}
\caption{A 60bp CDR3 read (grey box) can be aligned to different genes (nomenclature follows IMGT conventions \cite{Monod:2004im}) with different deletions (white), insertions (yellow), and P-nucleotides (red). (A) Alignment to specific V-, D-, and J-genes with insVD=$13$, insDJ=$6$, delV=$5$, delJ=$6$, del5'D=$6$, del3'D=$-2$ (in other words, pal3'D=$2$). (B) Alignment of the same read to different V- and D-genes, and with insVD=$15$, insDJ=$9$,delV=$7$, del5'D=$9$, del3'D=$3$ (no P-nucleotides). Note that the alignment to the V-gene is not maximal in this case. A few heavily penalized mismatches are allowed (in the V-gene in this example) in order to accommodate a small sequencing error rate. The location of the sequencing primer is indicated: it is chosen to uniquely identify the start of the CDR3 read within each J-gene.
\label{figcartoon}
}
\end{figure*}
}

% FIGURE 2

\newcommand{\paneltwo}{
\begin{figure}
\noindent\includegraphics[width=\linewidth]{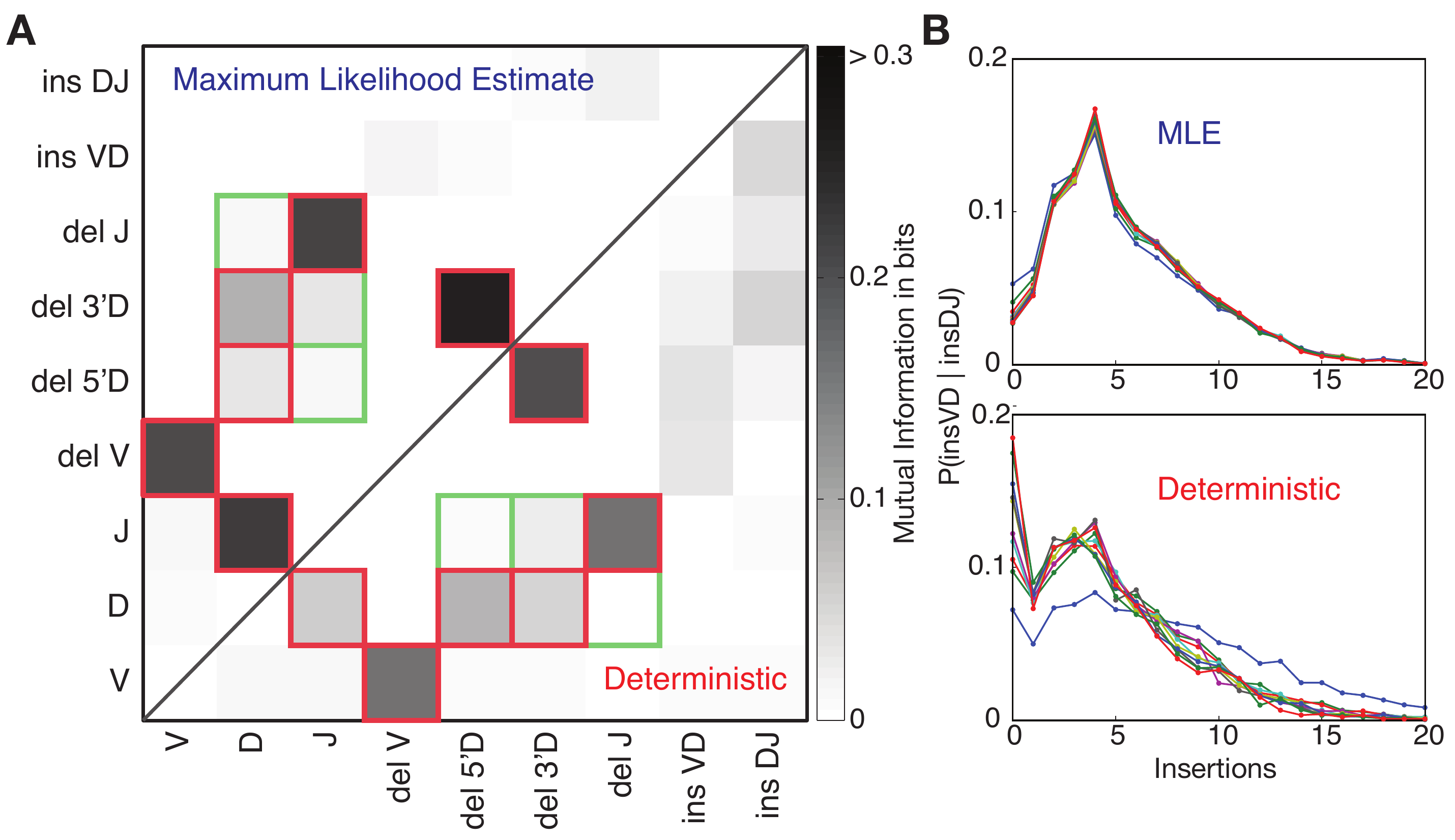}
\caption{
\label{figcorrelation} (A) Data-derived correlations between sequence
features: each entry is the mutual information $I(X,Y)$ of a feature
pair over the na\"ive non-productive repertoire. The outlined elements are correlations expected from the 
form of $P_{\rm recomb}(E)$: red identifies a direct effect of a
factor in Eqn.\,\ref{Precomb} (e.g. D $\leftrightarrow$ J) and green indirect effects (e.g. D $\leftrightarrow$ J $\leftrightarrow $ delJ). 
The top-left half of the matrix shows results from the maximum likelihood
estimate (MLE), while the bottom-right half corresponds to a deterministic maximum-alignment based identification of recombination events.
(B) Probability distribution of the number of
VD insertions conditioned on the number of DJ insertions for MLE
(top) and deterministic (bottom) analysis. Each curve corresponds to a different value of insDJ, ranging from 0 (blue) to 10. The curves collapse for MLE indicating independence.
}
\end{figure}
}

% FIGURE 4

\newcommand{\panelfour}{
\begin{figure}
\noindent\includegraphics[width=\linewidth]{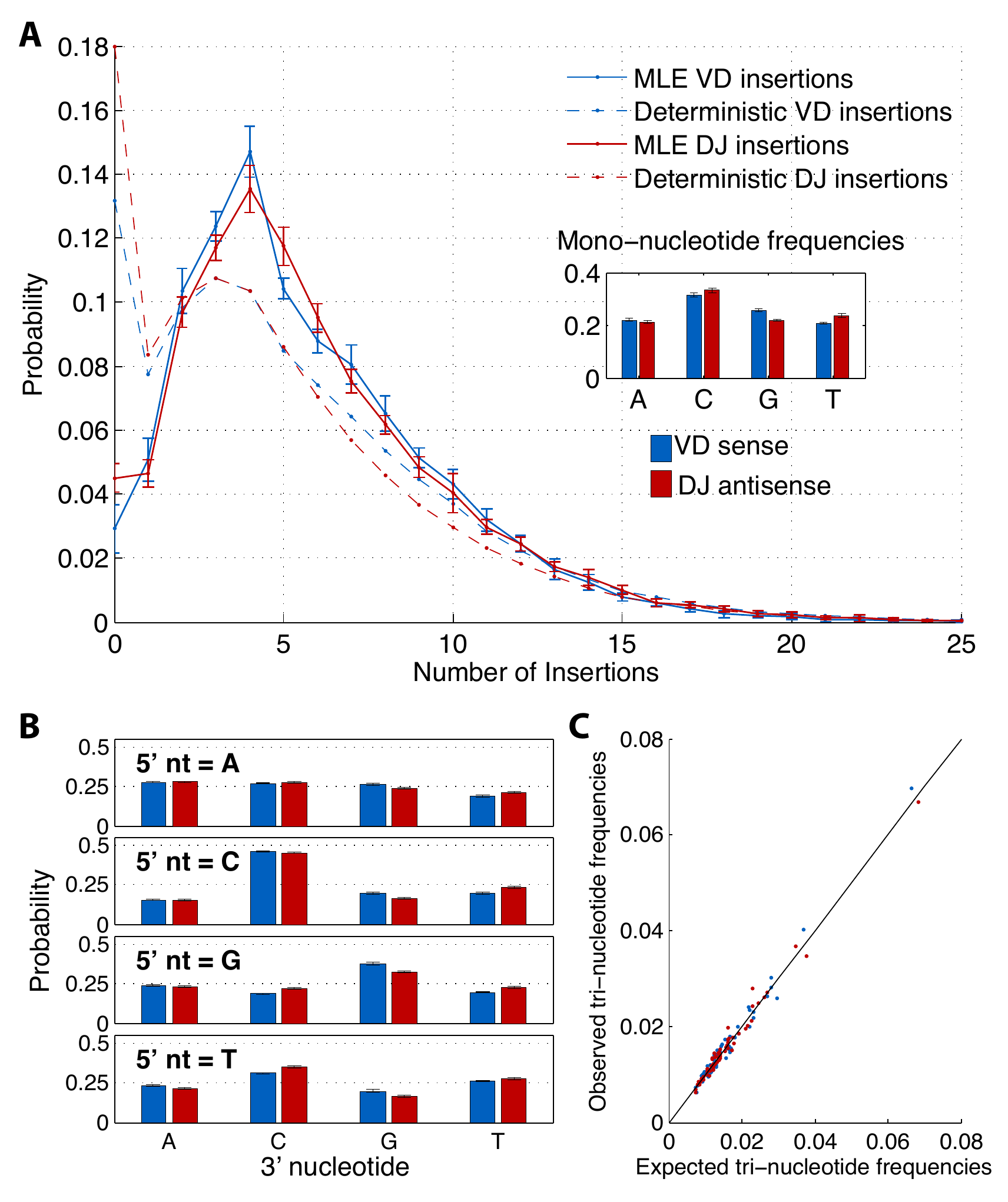}
\caption{Statistics of VD and DJ insertions. (A) Insertion length profiles: maximum likelihood estimate (deterministic estimate) displayed as solid (dashed) lines; error bars show variation across the nine individuals. The distribution tail is accurately exponential. The deterministic estimate greatly overestimates the frequency of zero insertions. Inset: mono-nucleotide utilization bias. (B) Dinucleotide utilization in insertions; the bias in DJ insertions is very accurately the reverse complement of the VD insertion bias.( C) Higher-order nucleotide bias in VD (blue) and DJ (red) insertions is completely accounted for by dinucleotide statistics.
\label{figinsertions}
}
\end{figure}
}

% FIGURE 5

\newcommand{\panelfive}{
\begin{figure}
\noindent\includegraphics[width=\linewidth]{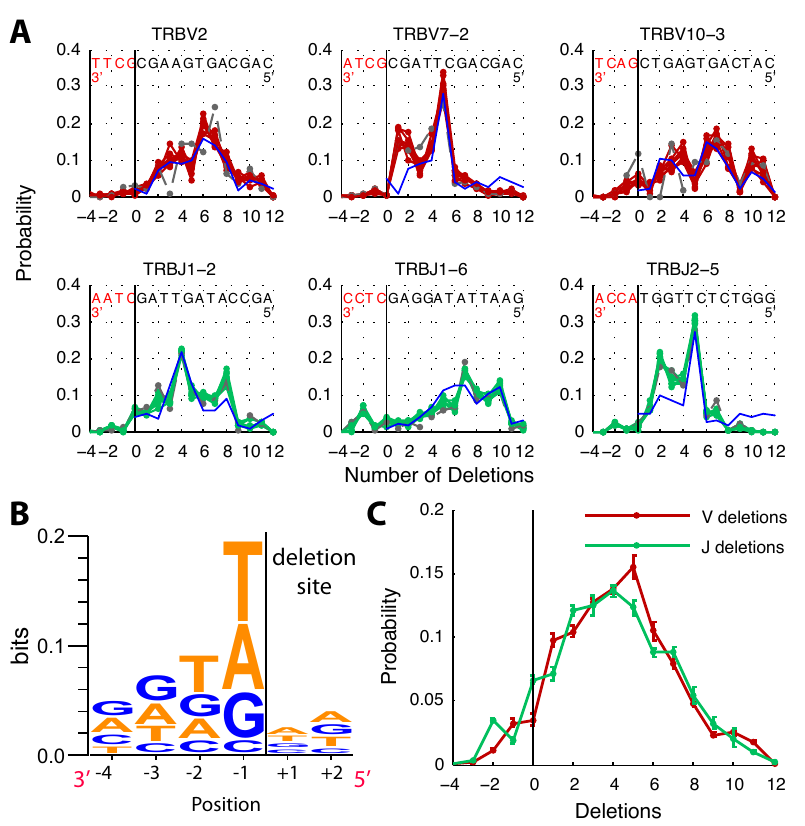}
\caption{(A) Gene-specific deletion profiles for selected V (red) and J (green) genes: the profiles vary widely from gene to gene, but are nearly identical across individuals (all nine are plotted; one in grey from an individual with significantly smaller sample size). The blue curves in all panels show the predictions of a simple model for the sequence context dependence of deletion probabilities using a position weight matrix (PWM), fit to the V deletion profiles (see Appendix \ref{delmodel} for details). The model ignores P-nucleotide generation and lacks any effects of distance from the gene end but performs reasonably well ($r^2 =0.7$). (B) Sequence logo of the context dependence of deletion probability, from the PWM fit to the V deletion profiles. Only positions \threep of the deletion site have strong effects on the probability. (C) Cumulative deletion profiles for V-genes and J-genes. Error bars indicate variation across individuals. 
\label{figdeletions}
}
\end{figure}
}

% FIGURE 6

\newcommand{\panelsix}{
\begin{figure}
\noindent\includegraphics[width=\linewidth]{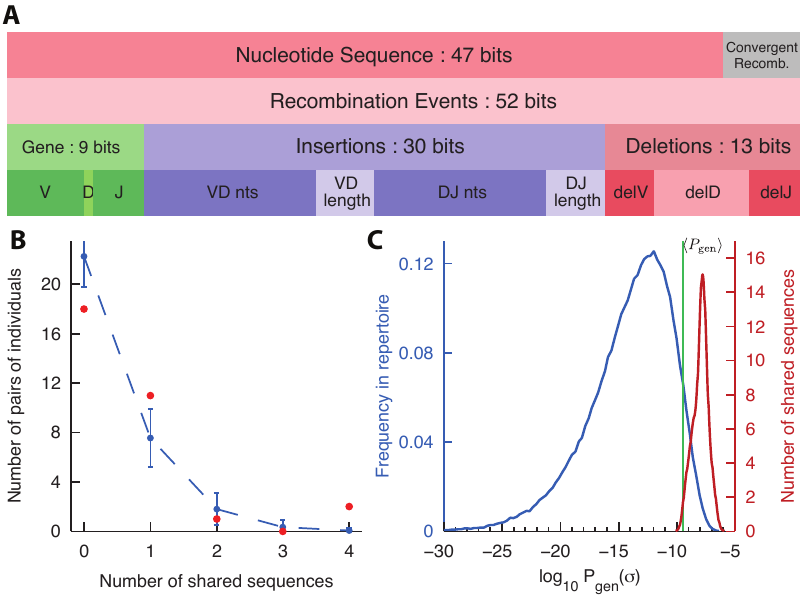}
\caption{ (A)  Entropy decomposition. Top bars: sequence entropy is smaller than recombination entropy by 5 bits because of convergent recombination; Bottom bars: recombination event entropy decomposed into contributions from gene choice, insertions, and deletions. (B) Statistics of the 21 CDR3 sequences shared between pairs of individuals: actual (red) vs. expected on the basis of the inferred $P_{\rm{gen}}(\sigma)$ (blue). (C) Histogram of $P_{\rm{gen}}(\sigma)$ for all sequences (blue) and for the 21 shared sequences (red, kernel density estimate); $\langle P_{\rm{gen}}\rangle$ for the full repertoire is indicated by the vertical green line.
\label{figrepertoire}
}
\end{figure}
}

\section{Introduction}

Receptor proteins on the surfaces of B- and T-cells in the immune system interact with pathogens, recognize them and initiate an immune response. The diversity of these receptors is the outcome of a remarkable process in which germline DNA is edited to produce a repertoire of (T- or B-) cells with varied antigen receptor genes \cite{Murphy:2008vw}. The process is called VDJ recombination because the germline contains multiple versions of so-called V-, D-, and J-genes, particular instances of which are quasi-randomly selected, stochastically edited and joined together to produce a new surface receptor gene each time a new immune system cell is generated.

The statistical distribution of these biochemical events (and the resulting receptor coding sequences) in a population of newly-created receptors is an important quantity: it contains information about the {\sl in vivo} functioning of the biochemical editing mechanism and provides the baseline for a quantitative assessment of the downstream workings of selection in the adaptive immune system. Here, we address the problem of inferring this distribution from the large T-cell sequence repertoires that are becoming available via high-throughput sequencing technology \cite{Robins:2009da, Robins:2010hda, Freeman:2009fja, Jiang:2011dt}. In particular, we focus purely on a subset of receptor sequences that are non-productive, due to a reading frame shift or an accidental stop codon, to isolate the statistics of the molecular mechanism from the effects of selection on the functional repertoires.

In the beta chain of human T-cell receptors (the focus of this work), the germline has 48 different V-genes, 2 D-genes and 13 J-genes. VDJ recombination proceeds by first joining a D-gene with a J-gene, and then a V-gene with the DJ junction.  First, the recombination activating gene (RAG) protein complex brings two randomly chosen D- and J-genes together, cuts out the intervening chromosomal DNA, and forms a hairpin loop at the end of each gene \cite{Swanson:2011vu,Verkaik:2002wv}. In further steps \cite{Lieber:2010ina,Lieber:2010dh} the hairpin loops are opened, creating overhangs at the end of both genes that may eventually survive as P-nucleotides (short inverted repeats of gene terminal sequence) \cite{Lafaille:1989td}. This is followed by nucleotide deletions and insertions at the junctions and ends with ligation. The process is then repeated between a random V-gene and the DJ junction. The end product is the so-called CDR3 region of the receptor gene: a short, highly variable region that plays an essential role in determining the antigen specificity of the cell.

Each recombined sequence can thus be thought of as the outcome of a generative event described by several random variables (Fig.\,\ref{figcartoon}): V-, D-, and J-gene choices, deletions of variable numbers of nucleotides from the selected genes, insertions of random nucleotides between them, and the possible creation of P-nucleotides (short palindromic nucleotides as in Fig.\,\ref{figcartoon}A at the \threep end of the D-gene). From the set of observed CDR3 sequences, we wish to infer the underlying probability distribution of these generative events. 

To date, this inference has been done via a deterministic alignment procedure which assigns a unique event to each sequence \cite{Freeman:2009fja, Jiang:2011dt}. 
However, since individual CDR3 sequences can arise in multiple ways (see Fig.\,\ref{figcartoon}), this assignment must be done probabilistically. Deterministic alignment introduces spurious biases and correlations in the statistics of generative events (Fig.\,\ref{figcorrelation}).
Thus, a statistical inference procedure is needed to accurately infer the underlying event probability distribution from the data. In this paper we present such a method, based on likelihood maximization via an iterative expectation-maximization algorithm \cite{McLachlan:2008wo},  and apply it to recent data on human T-cell receptor sequences. 

\section{Analysis strategy}

\panelone

We  work with sequence data on CD4+ T-cell beta chain CDR3 regions obtained from nine human subjects as described in \cite{Sherwood:2011ek}. In these experiments, T-cells are collected from a blood sample, and sorted into `na\"ive' (CD45RO-) and `memory' (CD45RO+) compartments, DNA is extracted, and sequence reads long enough to capture a \fivep piece of the J gene, a \threep piece of the V gene and the variable sequence lying in between, are obtained. Each sequence is read multiple times, and a clustering algorithm is used to correct for sequencing error \cite{Sherwood:2011ek}. This process produces a data set consisting of an average of 232,000 (140,000) unique CDR3 sequences from the na\"\i ve (memory) compartments for each individual subject \footnote{We are grateful to H. Robins and collaborators for making the data sets on which this work is based available to us.}. Each unique sequence comes with a multiplicity (ranging over three orders of magnitude) reflecting the prevalence of that particular cell type in the blood sample. 

Roughly 14\% of the unique CDR3 sequences are  \lq non-productive\rq\, \emph{i.e.} either their J genes have been shifted out of the correct reading frame or the CDR3 sequences have a premature stop codon. They arise from a recombination event on one of a cell's two chromosomes that failed to make a functional receptor, followed by a successful recombination on the other chromosome. Such sequences should not be subject to functional selection, and their statistics should reflect only the VDJ recombination process \footnote{We verify that the condition for being non-productive does not bias recombination event statistics by analyzing simulated sequences. See Appendix \ref{testsimul}.}. Since this is our primary concern, we focus our analysis on the non-productive CDR3 sequences, of which there are an average of 35,000 (22,000) in the na\"\i ve (memory) compartments for each individual subject. We analyze the na\"ive and memory data sets separately to be able to verify the absence of selection effects.

\subsection{Structure of recombination event distributions}
Each CDR3 generating recombination event can be fully characterized by a set $E$ of discrete variables comprising: the identities of the V-, D- and J-genes selected for recombination \footnote{Here we distinguish only the genes, not their various alleles. The gene list includes germline pseudo-genes: they cannot produce functioning receptor proteins but, since we work with non-coding VDJ rearrangements, pseudogene sequences can appear in the data.} (V,D,J); the numbers of bases deleted from the 3' end of the V-gene (${\rm del}V$), the 5' end of the J-gene (${\rm del}J$), and both ends of the D-gene (${\rm del} 5^{\,\prime} D$ and ${\rm del} 3^{\,\prime}D$ for the 5' and 3' ends, respectively); the number of palindromic nucleotides at each of the gene ends (${\rm pal}V, {\rm pal}J,{\rm pal}5^{\,\prime}D,{\rm pal} 3^{\,\prime}D$); the specific sequence $(x_1,\ldots,x_{{\rm ins}VD})$ of length ${\rm ins}VD$ inserted at the VD junction, and the specific sequence, $(y_1,\ldots,y_{{\rm ins}DJ})$ of length ${\rm ins}DJ$ inserted at the DJ junction (see Fig.\,\ref{figcartoon}). We choose a convention in which both sequences are read in the 5' to 3' direction, but the VD (DJ) inserted sequence is read from the sense (antisense) strand. 

We seek a joint distribution over all of these variables containing the minimal set of dependences between the variables that is required to self-consistently capture the observed correlations in the data. We find that the following factorized form for the probability of a recombination event $E$ (defined by specific values for all the event variables)  successfully captures all the significant correlations between sequence features that are present in the data (see Fig.\,\ref{figcorrelation}):

\beq\label{Precomb}
\begin{split}
&P_{\rm recomb}(E) = P(V) \, P(D,J) \times  \\
&\quad P({\rm del}V|V) \, P({\rm del}J|J) \, P({\rm del} 5^{\,\prime}D,{\rm del} 3^{\,\prime}D|D) \times  \\
&P({\rm ins}VD) \prod_{i=1}^{{\rm ins}VD} p^{(2)}_{VD}(x_{i}|x_{i-1})\,
%\\ &\times
P({\rm ins}DJ)\prod_{i=1}^{{\rm ins}DJ}p^{(2)}_{DJ}(y_{i}|y_{i-1}).
\end{split}
\eeq
The various factors are normalized joint or conditional distributions on their respective arguments. $P(V)$ and $P(D,J)$ account for the fact that the various genes have different usage probabilities (and that D- and J-gene usage is correlated). The factors $P({\rm del}V|V)$, etc., are distributions on the number of nucleotide deletions, conditioned on the gene being deleted (deletion profiles turn out to be very gene-dependent). $P({\rm ins}VD)$ and $P({\rm ins}DJ)$  give the probabilities of different numbers of nucleotide insertions at each junction. The parameters $p^{(2)}_{VD}$ and $p^{(2)}_{DJ}$ account for possible nucleotide bias in the insertions: they give the conditional probabilities of inserting a specific nucleotide given the identity of the immediately 5' nucleotide, with $x_{0}$ referring to the last nucleotide at the 3' end of the truncated V-gene on the sense strand for a VD insertion, or at the end of the truncated J-gene on the antisense strand for a DJ insertion.  

P-nucleotides do not appear explicitly in Eqn.\,\ref{Precomb}: we treat them as `negative' deletions  (\emph{i.e.} a palindrome of half-length 2, as in Fig.\,\ref{figcartoon}A, is counted as a deletion of value $-2$). This is possible because we find that when the number of nucleotide deletions is greater than zero, occurrences of palindromic nucleotides at the end of the gene segment are completely explained by chance insertions of the corresponding nucleotides (see  Fig.\,\ref{figpnuc}). Thus, true P-nucleotides, not attributable to chance insertions, only occur in association with zero nucleotide deletions and it is consistent to label them as `negative' deletions.

The factors in our equation for $P_{\rm recomb}(E)$  (Eqn.\,\ref{Precomb}) are probability distributions on event variables that take on a finite number of values. Specifying this joint distribution requires a total of 2865 probabilities (more than 90\% of which are needed for the  deletion length probabilities of the individual V-, D- and J-genes). Despite the large number of probabilities to be inferred, we are able to determine them accurately and without overfitting. We emphasize that our goal is to obtain an accurate description of recombination event statistics, and not (yet) to explain those statistics mechanistically.

\subsection{Generation probability and likelihood of observed sequences}
The probability $P_{\rm gen}(\sigma)$ of generating a specific CDR3 sequence $\sigma$ is the sum of the probabilities of all recombination events $E_{\sigma}$ that produce $\sigma$: 
\begin{align}
P_{\rm gen}(\sigma) = \sum_{E\in E_{\sigma}} P_{\rm recomb}(E) \label{Pgen}.
\end{align}
The likelihood $L(\sigma)$ of observing a specific CDR3 sequence read $\sigma$, however, must take into account residual sequencing error as well as allelic variation, and is given by a sum over a larger set of recombination events $\widetilde{E}_{\sigma}$ that generate sequences close to $\sigma$:
\begin{align}
&L(\sigma) = \sum_{E\in \widetilde{E}_{\sigma} } P(E,\sigma) \label{likelihood} \qquad {\rm where} \\
&P(E, \sigma) = P_{\rm recomb}(E)   \times \frac{1}{(1 + R)^{L}}\notag \\
&\times \sum_{{\rm alleles}\, a} P(V_{\rm a}|V_{E})P(J_{\rm a}|J_{E})P(D_{\rm a}|D_{E})\label{allele_error} \left(\frac{R}{3} \right)^{n_{\rm err}(\sigma_{E}^{a},\sigma)}.
\end{align}

\noindent In the latter equation, $n_{\rm err}$ is the number of mismatches between the observed read $\sigma$ and the CDR3 sequence $\sigma_{E}^{a}$ that would be produced by the recombination event $E$ with allele choices $a$. $L$ is the length of the sequence read. The mismatch rate $R$ is determined in the inference with the rest of the distribution parameters and reflects both sequencing error as well as unknown allelic variation. In practice, we only consider recombination events $\widetilde{E}_{\sigma}$ that lead to CDR3 sequences with at most a few mismatches from $\sigma$. The sum over alleles \footnote{We use the known alleles for each gene listed in the IMGT data base \cite{Monod:2004im} augmented by a few additional variants observed in the data (see Appendix \ref{alleles} for details).} arises because we do not know \emph{a priori} which alleles are present and reads may not go deep enough into the gene sequence to clearly distinguish alleles from each other \cite{Wang:2011fu}. The probabilities of the different alleles, given a gene, are also inferred and are expected to differ from individual to individual.

The likelihood of the whole data set $\mathcal{D}$ is then the product over the individual sequence likelihoods:
$\mathcal{L}(\mathcal{D})=\prod_{\sigma \in \mathcal{D}} L(\sigma)$.  This expression depends implicitly on the parameters defining the generative probability distribution (along with the allele distributions and the sequencing error parameter), and we infer their correct values by maximizing $\mathcal{L}(\mathcal{D})$ using an expectation maximization algorithm \cite{McLachlan:2008wo,Dempster:1977ul} (see Appendix \ref{overallanalysis} for details on the algorithm and its convergence). In order to identify universal features of the diversity generation machinery, we perform this inference separately for each individual subject.

\section{Results}
In what follows, we present results of our analysis of na\"ive, non-productive, CDR3 sequence repertoires of nine individuals (see Appendix \ref{memoryprod} for a parallel analysis of memory sequence repertoires). 

\subsection{Correlations between event variables}
It is important to verify that correlations not present in the assumed structure of the probability distribution  (Eqn.\,\ref{Precomb}) are in fact not present in the data. To perform this self-consistency check, we use the inferred generative distribution to compute the probability-weighted counts distribution of recombination event variables in the data, and then use this distribution to calculate the mutual information of all pairs of event variables. The matrix of mutual information values is shown in the upper-triangular part of Fig.\,\ref{figcorrelation}A, where the entries outlined in red are dependences accounted for by individual factors in our assumed form of $P_{\rm recomb}(E)$ (Eqn.\,\ref{Precomb}), entries outlined in green are indirect dependences that can be induced by these factors, and the rest would vanish if the data were perfectly described by the assumed structure of $P_{\rm recomb}(E)$. There are a few detectable correlations that are not consistent with the assumed structure: $(\mathrm{ins}VD, \mathrm{del}V), (\mathrm{ins}DJ, \mathrm{del}J)$ and $(\mathrm{V},\mathrm{D})$. They are, however, all so weak (mutual information $ < 0.02$ bits) that we do not model them explicitly (indeed, they might arise from subtle biases in our inference procedure).

For comparison, in the lower-triangular part of Fig.\,\ref{figcorrelation}A we show the mutual information values of all pairs of variables, but now calculated from a deterministic assignment of events to sequences based on maximal alignments. The resulting distributions exhibit spurious correlations that are absent from the corrected, maximum likelihood estimate (MLE) of the distributions. For instance, the number of insertions at the two junctions are found to be independent in our analysis while the uncorrected estimate shows a dependence (Fig.\,\ref{figcorrelation}B,C).

\paneltwo

\subsection{Gene usage distributions}
The inferred frequencies of V- and J-genes vary significantly from gene to gene, a phenomenon for which no mechanistic explanation has yet been given. In particular, linear location on the chromosome does not explain the pattern of either V- or J-gene usage (see  Fig.\,\ref{figgeneusage}A, C). The usage frequencies are consistent between individuals, though of all the inferred parameters in $P_{\rm recomb}$, these usage patterns show the most relative variation between individuals.

The pattern of D-gene use conditioned on J-gene choice (Fig.\,\ref{figgeneusage}D) reveals the known mechanistic constraint prohibiting utilization of D-genes that lie \threep of the chosen J-gene \cite{Murphy:2008vw}. The inferred distribution assigns a total probability of less than $0.1 \%$ for joining events using TRBD2 and any TRBJ1 gene. We note that such a determination is impossible without probabilistic analysis due to the uncertainty in identifying genes in specific sequences. The dependence between V gene choice and D or J gene choice is very weak to non-existent (with mutual information less than $0.01$ bits). Thus, we believe that previously reported correlations in the use of these genes \cite{Wallace:2000tl} reflect the effects of selection rather than VDJ recombination. Finally, we note the presence of pseudo V-genes which occur in almost $10\%$ of the non-productive CDR3s (see Appendix \ref{pseudogene} for more details).

\subsection{Nucleotide insertions}
In Fig.\,\ref{figinsertions} we show the factors related to insertions in the inferred distribution $P_{\rm recomb}(E)$. The VD and DJ insertions are uncorrelated (Fig.\,\ref{figcorrelation}) and their length distributions are nearly identical, with exponential tails (Fig.\,\ref{figinsertions}A). The nucleotide frequencies in the inserted segments are not uniform and are well explained by a di-nucleotide Markov model where the probability of inserting A, C, G, or T depends on the immediately 5' nucleotide (see Fig.\,\ref{figinsertions}B). The VD inserted segment, on the sense strand, and the DJ inserted segment, on the antisense strand, show a preference for Cs. The  frequencies of tri-nucleotides are almost perfectly accounted for by the di-nucleotide preferences (Fig.\,\ref{figinsertions}C), suggesting that the sequence statistics are fully captured by dinucleotide statistics. Additionally, the VD insertion di-nucleotide bias, taken on the sense strand in the 5'-3' direction, is virtually identical to the DJ insertion di-nucleotide bias, taken on the antisense strand in the 5'-3' direction. This suggests that the mechanism of junctional nucleotide insertions is strand specific and occurs on opposite strands for the VD and DJ junctions. The molecular mechanistic basis of these features is not evident.

\panelfour

\subsection{Nucleotide deletions}
Since there is a strong correlation between number of deletions and gene identity (see the entries for $I(\mathrm{del}V, V)$ and $I(\mathrm{del}J, J)$ in Fig.\,\ref{figcorrelation}), we allow for gene-dependent deletion profiles in $P_{\rm recomb}(E)$ (Eqn.\,\ref{Precomb}). The results for a few genes are shown in Fig.\,\ref{figdeletions}A (see Figs.\,\ref{figgenedeletions1}-\ref{figgenedeletions4} for all the profiles). P-nucleotides are counted as negative deletions as they occur only in association with zero nucleotide deletions (see Fig.\,\ref{figpnuc}). The profiles have substantial variation from gene to gene, suggestive of a nuclease activity that depends on sequence context, but they are highly consistent between individuals. We have modeled this context dependence using a position weight matrix summing independent contributions from the bases in a 6 nucleotide window (four \threep and two \fivep) around the cutting point to the log probability of deletion (see Fig.\,\ref{figdeletions}B and Figs.\,\ref{figgenedeletions1}-\ref{figgenedeletions4} for details). We find that only bases \threep of the deletion site have a strong effect on the probability, with T and A nucleotides having the greatest contribution, consistent with previous observations \cite{Gauss:1996vxa}. This simple model, which ignores both the P-nucleotides as well as the effects of distance from the end of the gene, does reasonably well in explaining the variation in deletion probabilities ($r^{2}=0.7$). This modeling is simply to suggest that the complexity of the observed  deletion distributions may ultimately be explained by a parsimonious mechanistic model that reflects the underlying biochemistry of the deletion process. 

\panelfive

\subsection{Consistency of distributions across individuals}
The insertion profiles, and the many different gene-dependent deletion profiles, are very consistent between individuals (Figs.\,\ref{figinsertions}, \ref{figdeletions} and Figs.\,\ref{figgenedeletions1}-\ref{figgenedeletions4}), suggesting the action of a universal molecular mechanism of rearrangement and providing convincing evidence against overfitting. We note that finite sample size statistics accounts for less than $50\%$ of the observed inter-individual variance (indicated by the error bars) in some of our plots, possibly reflecting biological variation.

\subsection{Potential diversity of repertoire}
Our inferred distribution of recombination events (Eqn.\,\ref{Precomb}) implies a probability distribution $P_{\rm gen}(\sigma)$ on the space of all CDR3 sequences (Eqn.\,\ref{Pgen}) whose entropy $S_{\rm seq}=-\sum_{\sigma}P_{\rm gen}(\sigma)\log P_{\rm gen}(\sigma)$ 
is a measure of the potential sequence diversity of VDJ recombination. Since multiple recombination events can lead to the same sequence, we cannot calculate $S_{\rm seq}$ directly. We do, however, have an explicit description of $P_{\rm recomb}$, the entropy of which we can calculate: $S_{\rm recomb}=52$ bits; in addition, we can show that sequence entropy and recombination event entropy are related by  
\beq\label{eq:S}
S_{\rm seq}=S_{\rm recomb}- \langle S(E|\sigma) \rangle_{\sigma}\simeq 47\textrm{ bits}\, ,
\eeq
where the correction term, $\langle S(E|\sigma) \rangle_{\sigma}\simeq5$ bits, is the entropy of recombination events that give the same sequence, averaged over sequences.
This means that CDR3 sequences can be generated in $\sim 32$ different ways, on average, by VDJ recombination; this is the fundamental reason why we must resort to probabilistic inference methods. The total sequence diversity of $47$ bits corresponds to a potential CDR3 repertoire size of $\sim 10^{14}$ sequences \footnote{Recall that this estimate is for the $\beta$-chain only. The $\alpha$-chain will yet add more diversity to this estimate.}. This is to be compared with the estimated $4 \times 10^{6}$  unique CDR3 sequences in an individual \cite{Arstila:1999uu, Robins:2009da} , the $\sim 10^{11}$  T-cells in the blood of an individual \cite{Blum:2007tn} and the $\sim 10^{13}$ potential peptide-MHC complexes \cite{Mason:1998ug}. While convergent recombination means that the sequence entropy cannot be neatly partitioned into contributions from gene choice, deletions and insertions, the entropy of recombination events $S_{\rm recomb}$ can be so partitioned (Fig.\,\ref{figrepertoire}A). We note that the bulk (60\%) of the recombination entropy comes from the nucleotide insertions, and little from gene choice ($5$ bits from V and $4$ bits from D and J) consistent with previous estimates \cite{Cabaniols:2001us}. For comparison, uniform usage of the genes would result in an entropy of $5.9$ bits for V and $4.7$ bits for D and J gene choices.

\panelsix

\subsection{Overlap of repertoires between individuals}
Some sequences appear in the repertoires of more than one individual, and we can ask whether their number and specific identities are consistent with chance on the basis of our generative distribution $P_{\rm gen}(\sigma)$. We see evidence of inter-sample contamination in some of our data leading to a large number of shared sequences between specific individuals. Eliminating such questionable cases (see SI Appendix for details), we are left with 21 sequences that occur in the non-productive repertoires of two individuals and none that occur in more than two.

The total number of shared sequences between the repertoire samples of any pair of individuals with sample sizes $N_{1}$ and $N_{2}$ is expected to be Poisson distributed with mean $\bar{n} = N_{1}N_{2} \langle P_{\rm{gen}} \rangle_{\sigma}$ where $\langle P_{\rm{gen}} \rangle_{\sigma} = \sum_{\sigma} P_{\rm gen}^{2}(\sigma)$. Note that while the specific shared sequences are likely to have high probabilities of generation, the number of shared sequences, without regard to their identities, is determined by $\langle P_{\rm{gen}} \rangle_{\sigma}$ which is the average value of $P_{\rm gen}$ over the potential repertoire. We estimate this quantity to be $\langle P_{\rm{gen}} \rangle_{\sigma} \simeq 3.4 \pm 0.1 \times 10^{-10}$, by taking the mean of $P_{\rm gen}$ over the observed repertoire.

In Fig.\,\ref{figrepertoire}B, we compare the expected number of pairs of individuals with a certain number of shared sequences (calculated as a sum of Poisson distributions over the pairs) to the observed number of such pairs, showing excellent agreement. The specific shared sequences have particularly high generation probabilities according to our distribution, with a median value of $\sim 10^{-8}$ compared to the repertoire median of $\sim 10^{-14}$ (Fig.\,\ref{figrepertoire}C). Since the generative distribution is trained on individual repertoires, and is highly consistent between individuals, its success in accounting for recurring sequences between individuals is a non-trivial test of its validity. We find similar results for the shared sequences among the memory repertoires (see Fig.\,\ref{figmemoverlap}). 

Convergent recombination has been proposed as an explanation for the occurrence of `public' TCRs \cite{Quigley:2010ff,Venturi:2011co,Venturi:2008hya}. However, the recombination entropy $S(E|\sigma)$ is only weakly correlated with the generation probability $P_{\rm gen}(\sigma)$ (correlation coefficient $0.13$, see Fig.\,\ref{figentropyvspgen}), and we find that the shared non-productive sequences in our data do not have higher recombination entropies than other sequences. 

\subsection{Results from other repertoires}
Inference of $P_{\rm recomb}(E)$ from the non-productive memory repertoires of the same nine individuals leads to results identical with those reported above for the na\"ive non-productive repertoires (see Fig.\,\ref{fignaivevsmem},\ref{figentropyvspgen}). The consistency of the inferred generative distribution between these repertoires as well as between the nine individuals is strong evidence that the non-productive CDR3 sequence statistics, memory or na\"ive, reflect only the basic recombination process and not selection. In Fig.\,\ref{figproductivePgen}, we show the distributions of generation probabilities of CDR3 sequences from the productive repertoires. While it is tempting to apply our algorithm to the productive sequence repertoires, it would be inconsistent to do so: these sequences have passed selection filters, thymic and adaptive, and we have no analog of Eqn.\,\ref{Precomb} to parametrize the probability of such success. This is an important subject for future investigation. 

\section{Discussion}

We have presented a method for inferring the statistics of VDJ recombination events from the large T-cell receptor sequence repertoires that are now available by high-throughput sequencing. We emphasize the crucial importance of using a probabilistic approach: the typical CDR3 sequence can be produced by about $32$ different recombination events, and using a deterministic assignment of events to each sequence results in systematic biases and spurious correlations. Our general approach allows us to cope with not-yet-indexed alleles \cite{Wang:2011fu} and, most importantly, with sequencing errors, an essential task given the rapid growth of high-throughput but error-prone sequencing technologies.

Since we focus on non-productive sequences, our results describe the probability distribution over CDR3 sequences produced by the recombination machinery {\em before any functional selection has occurred}. Its remarkable reproducibility across individuals and repertoires (na\"\i ve and memory) provides compelling evidence for the consistency and accuracy of our method.
The obtained distribution is a central feature of the adaptive immune system and serves as a baseline (or, in evolutionary terms, a neutral model) for analyzing the subsequent processes of the immune system. By calculating the entropy of the generative distribution, we can estimate the potential diversity of the CDR3 sequences ($\sim 10^{14}$ sequences) and the contributions of insertions, deletions and gene choices to this entropy. We find that insertions contribute most (60\%) of the diversity.

We are able to evaluate the probability of generating {\em any} specific CDR3 sequence (including as yet unobserved ones). This probability could be used to estimate the strength of selection on a sequence or group of sequences, or the likelihood that a sequence is shared between individuals or repertoires. Thus, it could help better characterize the significance of shared or `public' TCR sequences \cite{Venturi:2008hya}. We have verified that the sequences that are shared between the non-productive repertoires of different individuals in our data are consistent with the predictions of the inferred probability distribution (Fig.\,\ref{figrepertoire}B,C), a very stringent test of its accuracy.

The recombination event distributions also provide insight into the molecular mechanism of recombination, and should serve as a starting point for detailed mechanistic models of recombination. We find that the recombination processes at the two junctions are essentially independent of each other, and that insertion events are independent of gene choice and deletions. The inferred distribution confirms that a D-gene can only recombine with downstream J-genes. We derive a precise model for the composition of inserted nucleotides, based solely on frequencies of di-nucleotides. We also show that a relatively crude model of sequence-specific nuclease activity can account for the deletion probabilities reasonably well. Our observed distribution, which is specified by a large number of probabilities, should be reproduced by parsimonious, but more realistic, mechanistic models.

We have focused on characterizing the molecular generation of nucleotide sequences that code for T-cell receptors. The functional receptor repertoire is first shaped by this molecular process and then by thymic selection and adaptation to pathogens. Quantitative models of the latter processes are needed for understanding the adaptive immune system. While the underlying biochemistry conveniently served to parametrize our sequence distributions, finding an analogous functionally relevant parametrization of amino-acid sequences to model the effects of selection is much more challenging \cite{Mora:2010jxa}.  Statistical analysis of the productive receptor repertoires, with our precise characterization of the unselected repertoire in hand, will hopefully aid in this effort.

\begin{acknowledgments}
{The work of CGC was supported in part by National Science Foundation grant PHY-0957573 and by US Department of Energy grant DE-FG02-91ER40671. The work of AM was supported in part by the NSF Physics of Living Systems program (PHY-1022140). CGC thanks the Institute for Advanced Study for hospitality during the performance of part of this work. The authors declare no conflicts of interest. }
\end{acknowledgments}

% FIGURE 3

\newcommand{\panelgeneusage}{
\begin{figure*}[htbp]
\noindent\includegraphics[width=.75\linewidth]{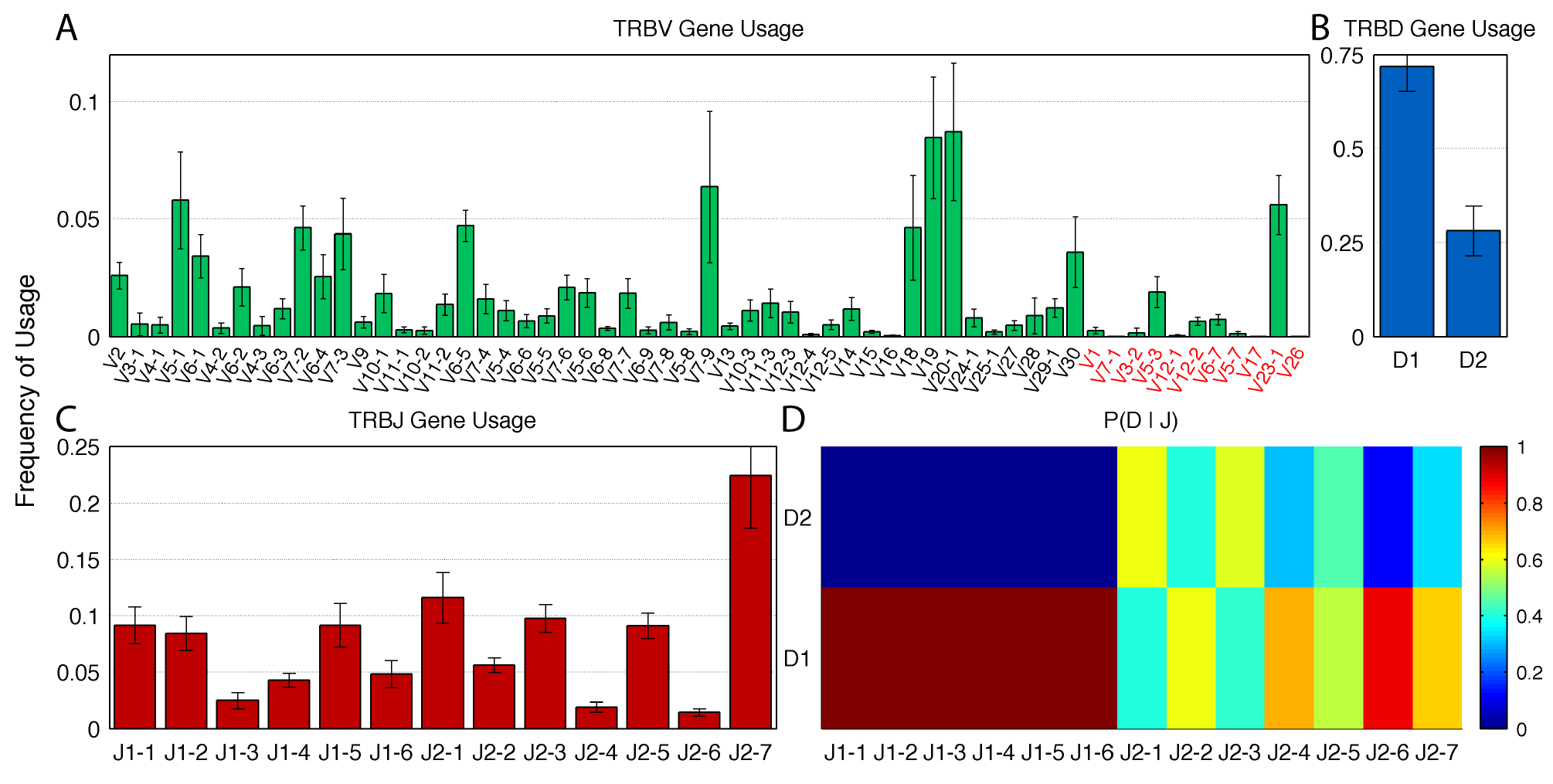}
\caption{Statistical aspects of gene usage. (A) Usage frequencies of V-genes, ordered by position on the chromosome, with the exception of pseudogenes (red legend). (B) Usage frequencies of the two D-genes. (C) Same for the 13 J-genes. (D) D-gene usage frequencies, conditioned on J-gene choice. As expected from the mechanistic constraint, TRBD2 has essentially zero probability ( $< 0.1\%$) of recombining with any TRBJ1 gene. Error bars indicate variation across the nine individuals.
\label{figgeneusage}
}
\end{figure*}
}

% FIGURE ? (Calculation Flow Chart)

\newcommand{\panelflowchart}{
\begin{figure}[htbp]
\noindent\includegraphics[width=\linewidth]{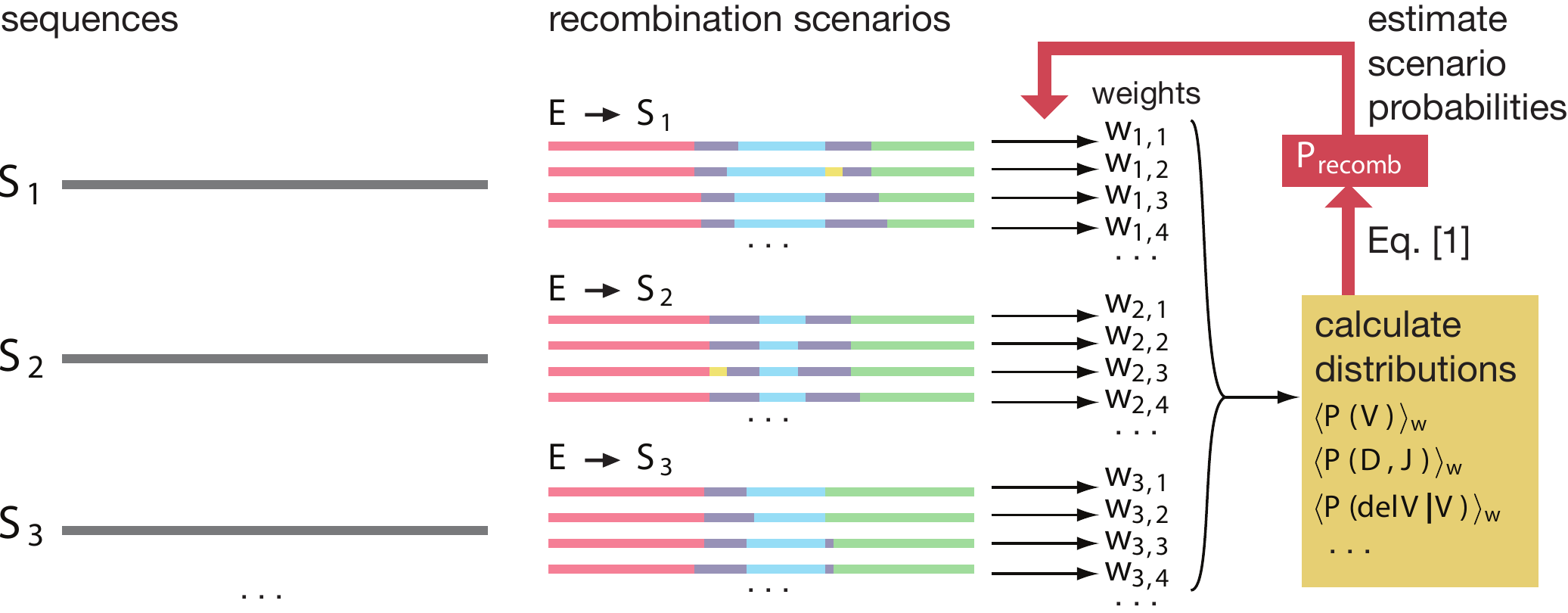}
\caption{Flow chart of the analysis pipeline.
\label{figflowchart}
}
\end{figure}
}

% FIGURE ? (Likelihood)
\newcommand{\panellikelihood}{
\begin{figure}[htbp]
\begin{center}
\noindent\includegraphics[width=\linewidth]{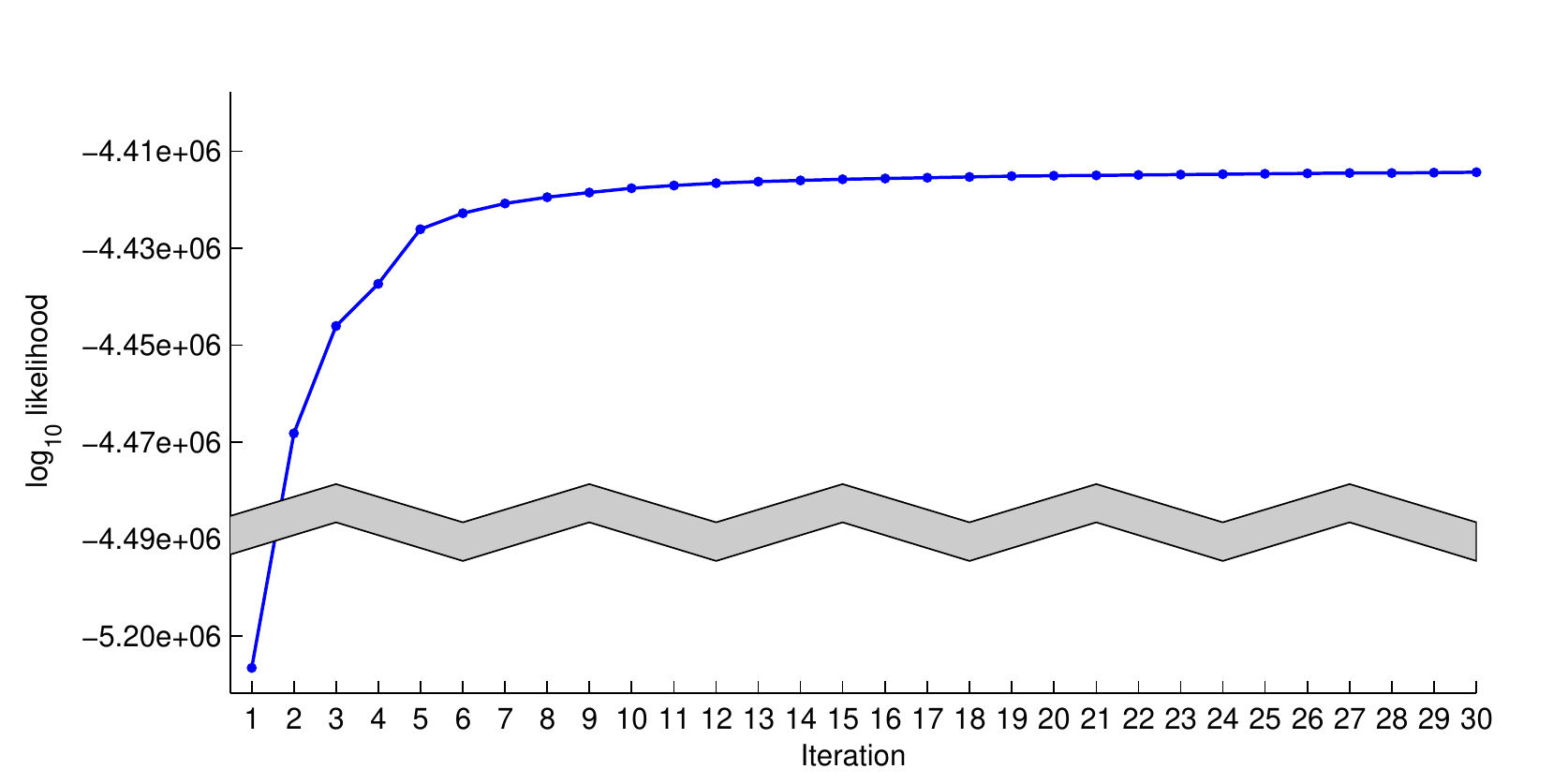}
\end{center}
\caption{Convergence of the total likelihood of all data sets with iterations of the EM algorithm.
\label{figlikelihood}
}
\end{figure}
}

%
%% FIGURE ? (Naive vs Mem)
\newcommand{\panelnaivevsmem}{
\begin{figure*}[htbp]
\begin{center}
\noindent\includegraphics[width=0.75\linewidth]{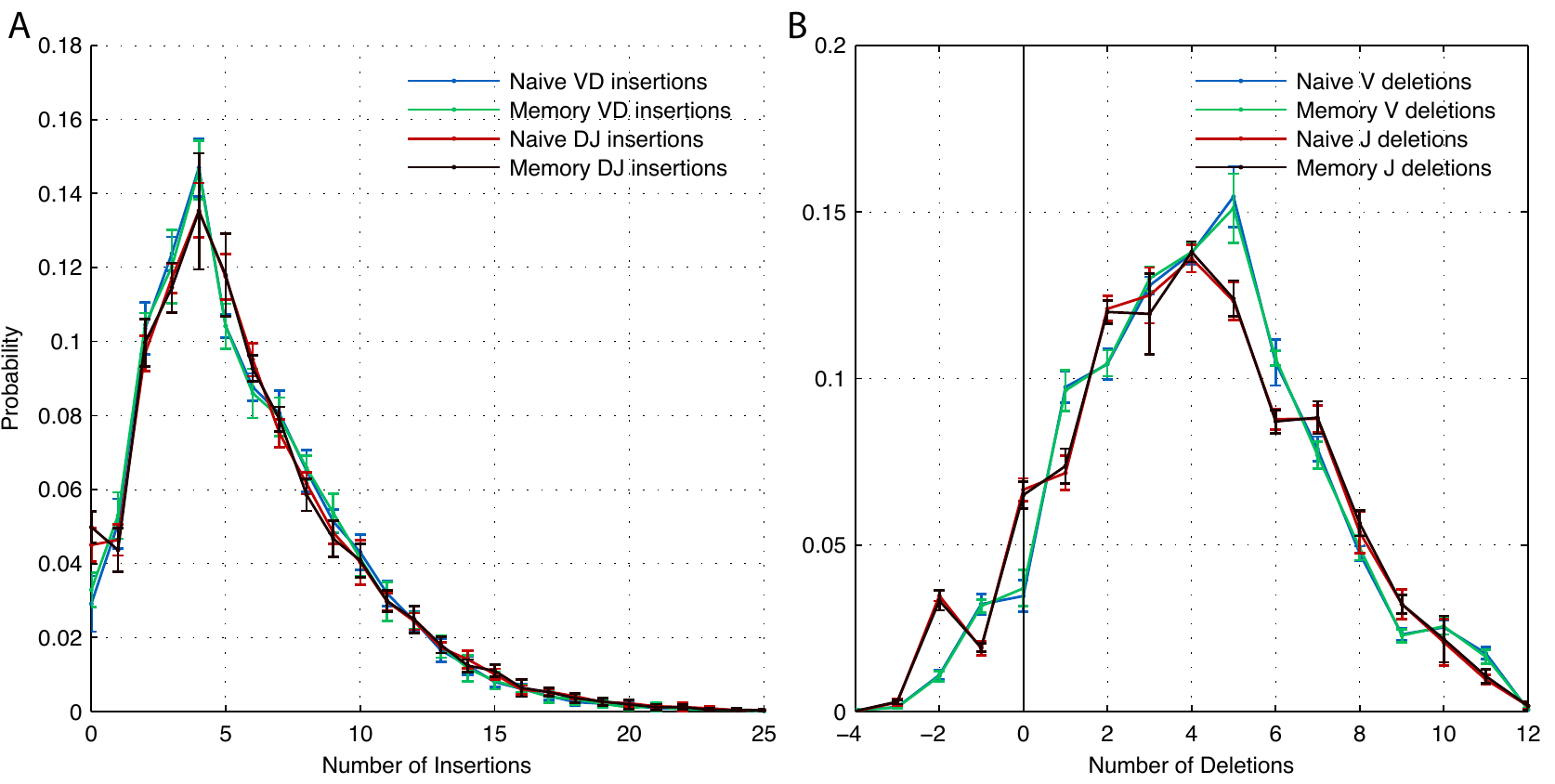}
\end{center}
\caption{Comparison of insertions (A) and deletions (B) distributions for the naive and memory T-cell repertoires. We find that the inferred models from the two compartments are statistically identical in all respects. Error bars indicate variation across the nine individuals.
\label{fignaivevsmem}
}
\end{figure*}
\begin{figure*}[htbp]
\begin{center}
\noindent\includegraphics[width=0.75\linewidth]{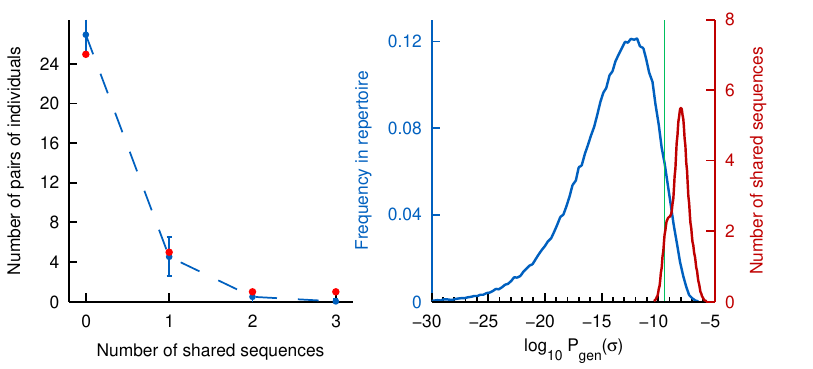}
\end{center}
\caption{Shared sequences in memory T-cell non-productive CDR3 sequence repertoires. A) Distribution of number of shared sequences between the 9 individuals. B) Distribution of $P_{\rm gen}(\sigma)$ for the entire repertoire (blue) and for the recurring sequences (red). $\langle P_{\rm gen} \rangle$ is indicated by the green vertical line.
\label{figmemoverlap}
}
\end{figure*}
}

% FIGURE ? (Entropy vs Pgen)
\newcommand{\panelentropyvspgen}{
\begin{figure}[htbp]
\begin{center}
\noindent\includegraphics[width=\linewidth]{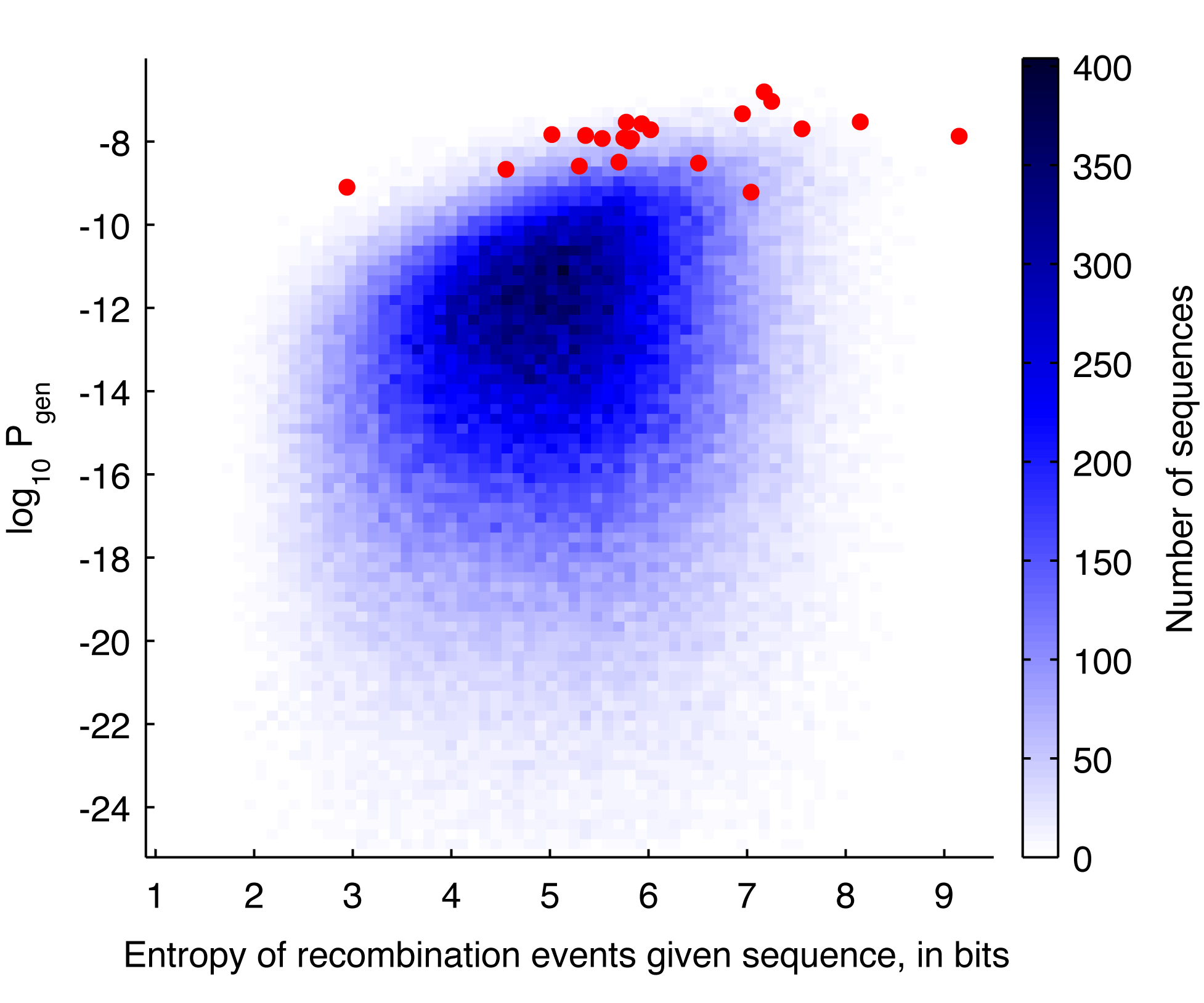}
\end{center}
\caption{A 2D histogram of conditional entropy of recombination events given the sequence and $P_{\rm gen}(\sigma)$. Convergent recombination (as measureed by the recombination event entropy) is a contributing factor to $P_{\rm gen}(\sigma)$, with correlation coefficient $0.13$. The shared sequences in the naive non-productive repertoires are shown in red.
\label{figentropyvspgen}
}
\end{figure}
}

% FIGURE ? (Error Rate Profile)

\newcommand{\panelerrorprofile}{
\begin{figure}[htbp]
\begin{center}
\noindent\includegraphics[width=\linewidth]{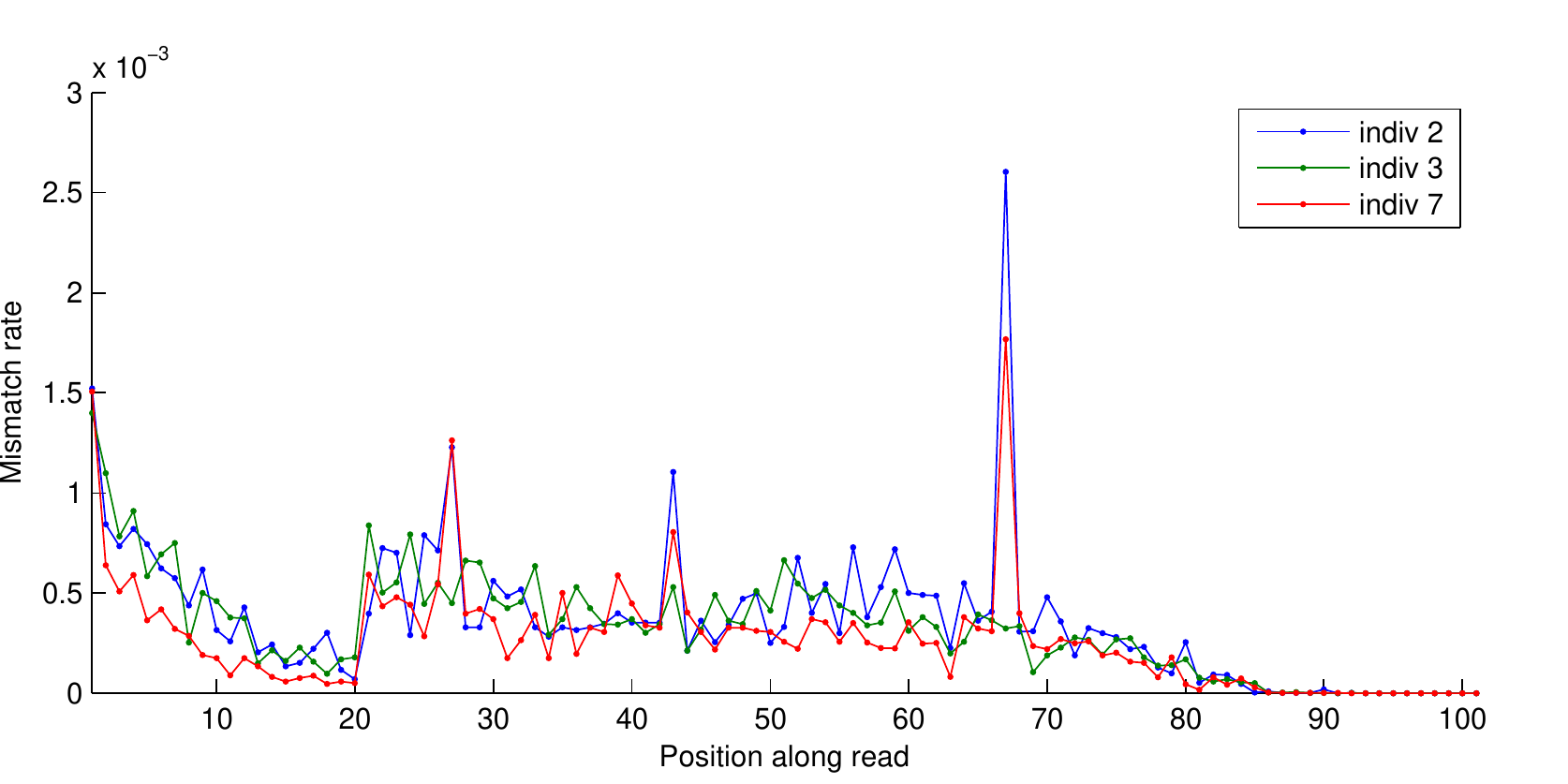}
\end{center}
\caption{Position-dependent error profile for the three individuals with read length 101 base pairs. The sequencing read proceeds from the right (101 to 1) where the J gene sequencing primer binds. The spikes in the error rate at specific positions (67, 43 and 27) are true sequencing error spikes and not the result of unknown allelic variants. Positions 1-15 show the characteristic increase in error rate with read length. The overall decreased error rate in positions 10-20 reflect our requirement of a minimum alignment length of 20 nucleotides to a V gene with an upper bound on the allowed errors in the alignment. Since we do not allow any errors in the J and D genes, the error rate is zero in this region.
\label{figerrorprofile}
}
\end{figure}
}

% FIGURE ? simulated seqs

\newcommand{\panelsimulated}{
\begin{figure*}[htbp]
\begin{center}
\noindent\includegraphics[width=\linewidth]{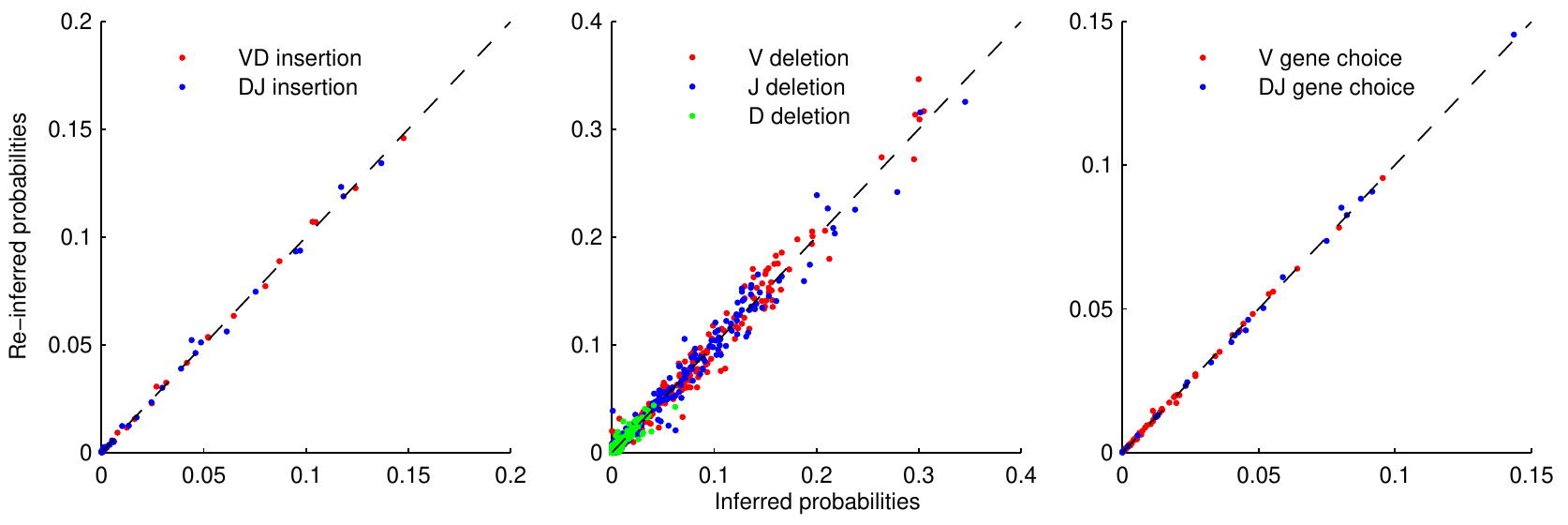}
\end{center}
\caption{Probabilities of recombination event variables were re-inferred by simulating sequences from our final distributions, discarding all in-frame sequences, and running the expectation-maximization algorithm on the out-of-frame subset. The above scatter plots show that the original probabilities are obtained. This provides evidence that the use of just the non-productive TCR sequences does not bias the statistics of recombination events.
\label{figsimulated}
}
\end{figure*}
}

\newcommand{\panelproductive}{
\begin{figure}[htbp]
\begin{center}
\noindent\includegraphics[width=\linewidth]{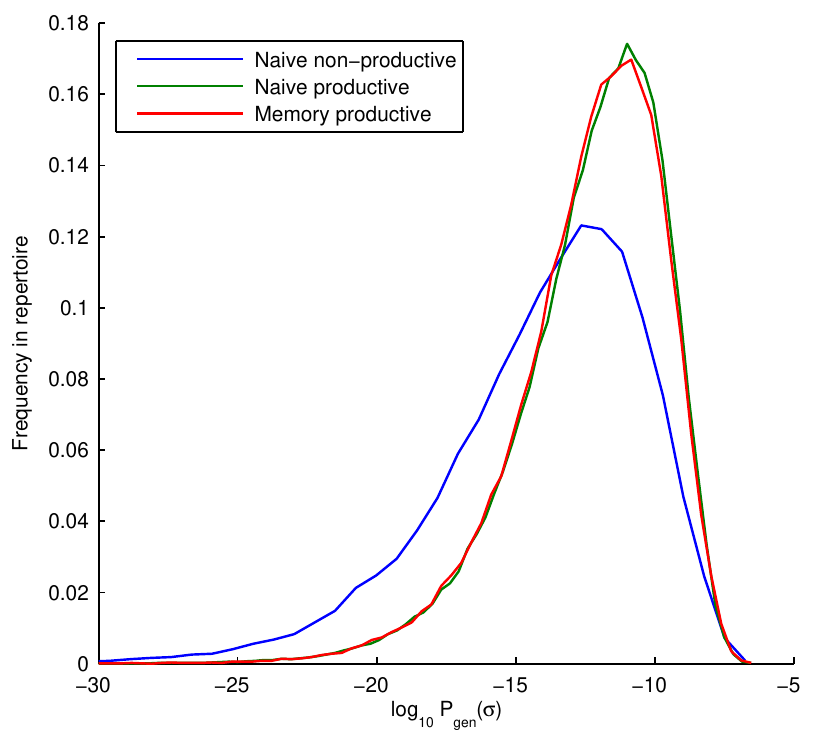}
\end{center}
\caption{Generation probabilities of all the CDR3 sequences in the naive and memory productive repertoires were computed using our inferred distribution. The above panel shows the distribution of the logarithm of these probabilities for the three repertoires for one individual. The productive repertoires have systematically higher generation probabilities.
\label{figproductivePgen}
}
\end{figure}
}

% FIGURE ? (P-nucleotide story)

\newcommand{\panelpnuc}{
\begin{figure}[htbp!]
\begin{center}
\noindent\includegraphics[width=\linewidth]{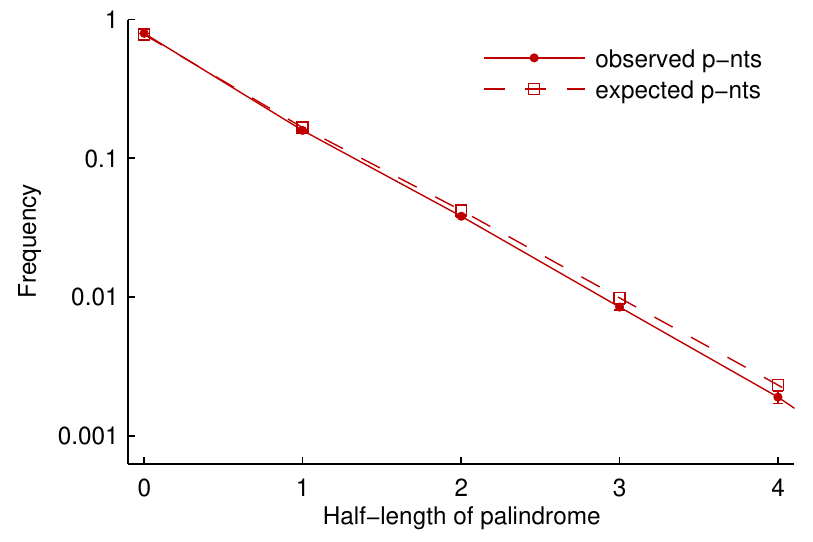}
\end{center}
\caption{Occurrence frequency of P-nucleotides for non-zero deletions.
\label{figpnuc}
}
\end{figure}
}

% FIGURE ? (Deletions bias story)
\newcommand{\paneldelseqmotif}{
\begin{figure}[htbp!]
\begin{center}
\noindent\includegraphics[width=\linewidth]{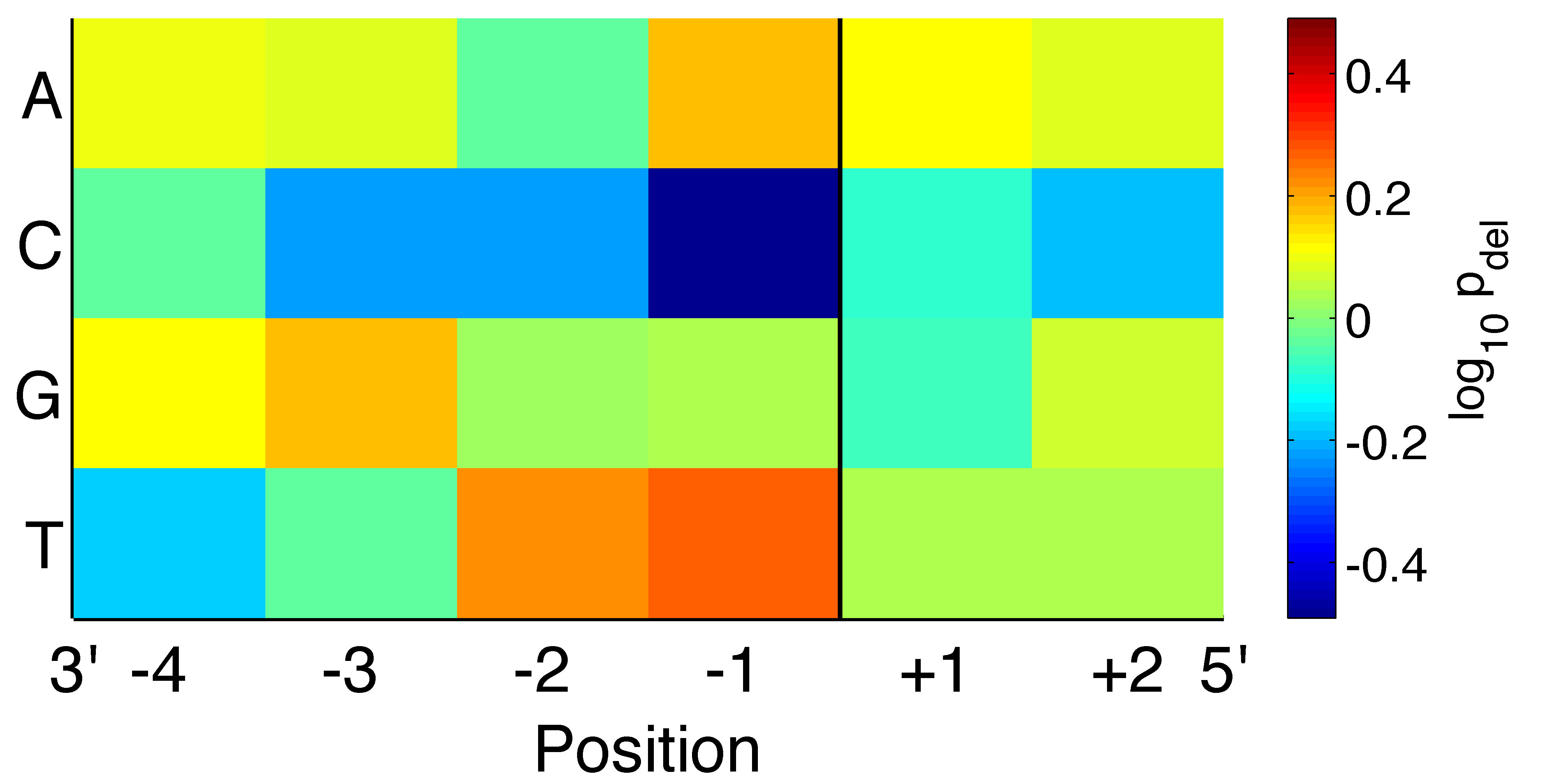}
\end{center}
\caption{Position weight matrix for sequence dependence of nucleotide deletion position. The figure shows $\epsilon/ \log(10)$ (see text of Appendix L) fit to the V gene specific deletions profiles, using four nucleotides \threep and two nucleotides \fivep of the deletion position (black vertical line). The \threep nucleotides are the most informative about deletion probability and show a preference for T and A. The sequence logo corresponding to this position weight matrix is shown in the main text Fig.\,\ref{figdeletions}B.
\label{figdelseqmotif}
}
\end{figure}
}

% FIGURE ? (V deletions panels)

\newcommand{\panelgenedeletions}{

\begin{figure*}[htbp!]
\begin{center}
\noindent\includegraphics[width=\linewidth]{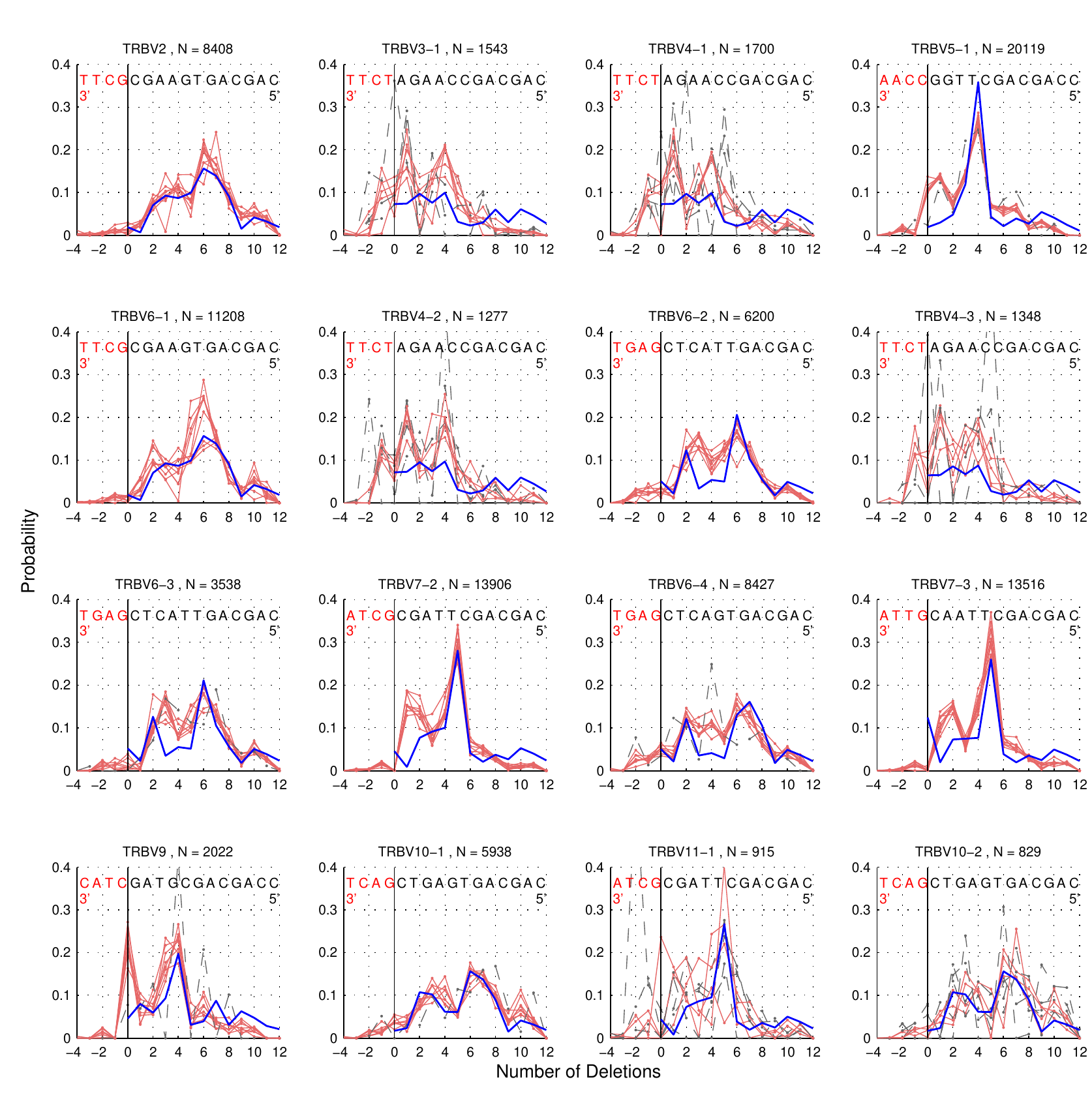}
\end{center}
\caption{Deletion profiles for all the V-genes (1 of 3). The title for each panel lists the gene name and total number of counts, across all the individuals studied, of the particular gene in question. Individuals with fewer than 100 counts for a specific gene are plotted in gray dashed lines. The blue lines show the predictions of the position weight matrix based model fit to these curves.
\label{figgenedeletions1}
}
\end{figure*}
~
\newpage
~
\newpage

\begin{figure*}[htbp!]
\begin{center}
\noindent\includegraphics[width=\linewidth]{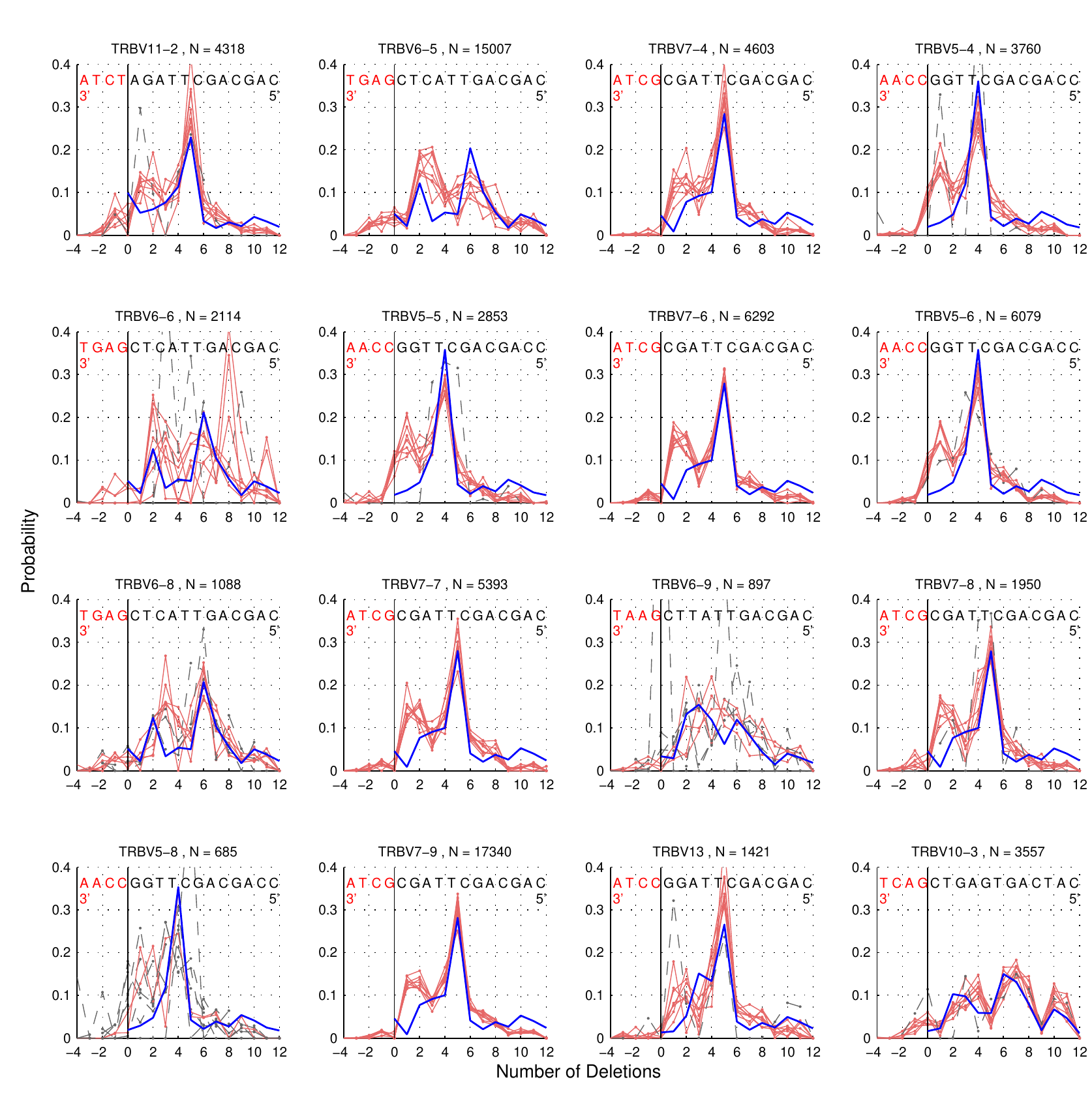}
\end{center}
\caption{Deletion profiles for all the V-genes (2 of 3). The title for each panel lists the gene name and total number of counts, across all the individuals studied, of the particular gene in question. Individuals with fewer than 100 counts for a specific gene are plotted in gray dashed lines. The blue lines show the predictions of the position weight matrix based model fit to these curves.
\label{figgenedeletions}
}
\end{figure*}
~
\newpage
~
\newpage

\begin{figure*}[htbp!]
\begin{center}
\noindent\includegraphics[width=\linewidth]{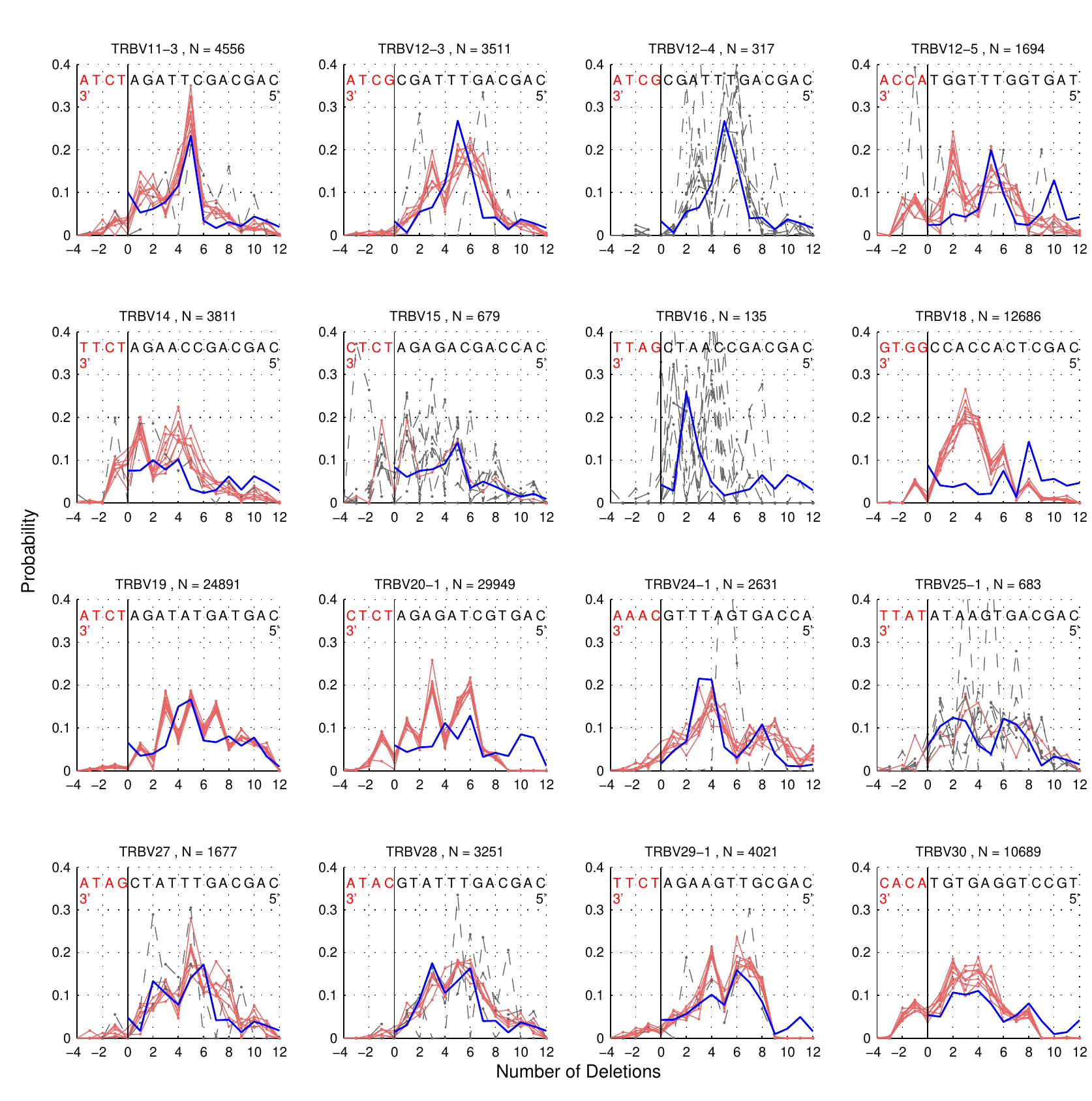}
\end{center}
\caption{Deletion profiles for all the V-genes (3 of 3). The title for each panel lists the gene name and total number of counts, across all the individuals studied, of the particular gene in question. Individuals with fewer than 100 counts for a specific gene are plotted in gray dashed lines. The blue lines show the predictions of the position weight matrix based model fit to these curves.
\label{figgenedeletions}
}
\end{figure*}
~
\newpage
~
\newpage

\begin{figure*}[htbp!]
\begin{center}
\noindent\includegraphics[width=\linewidth]{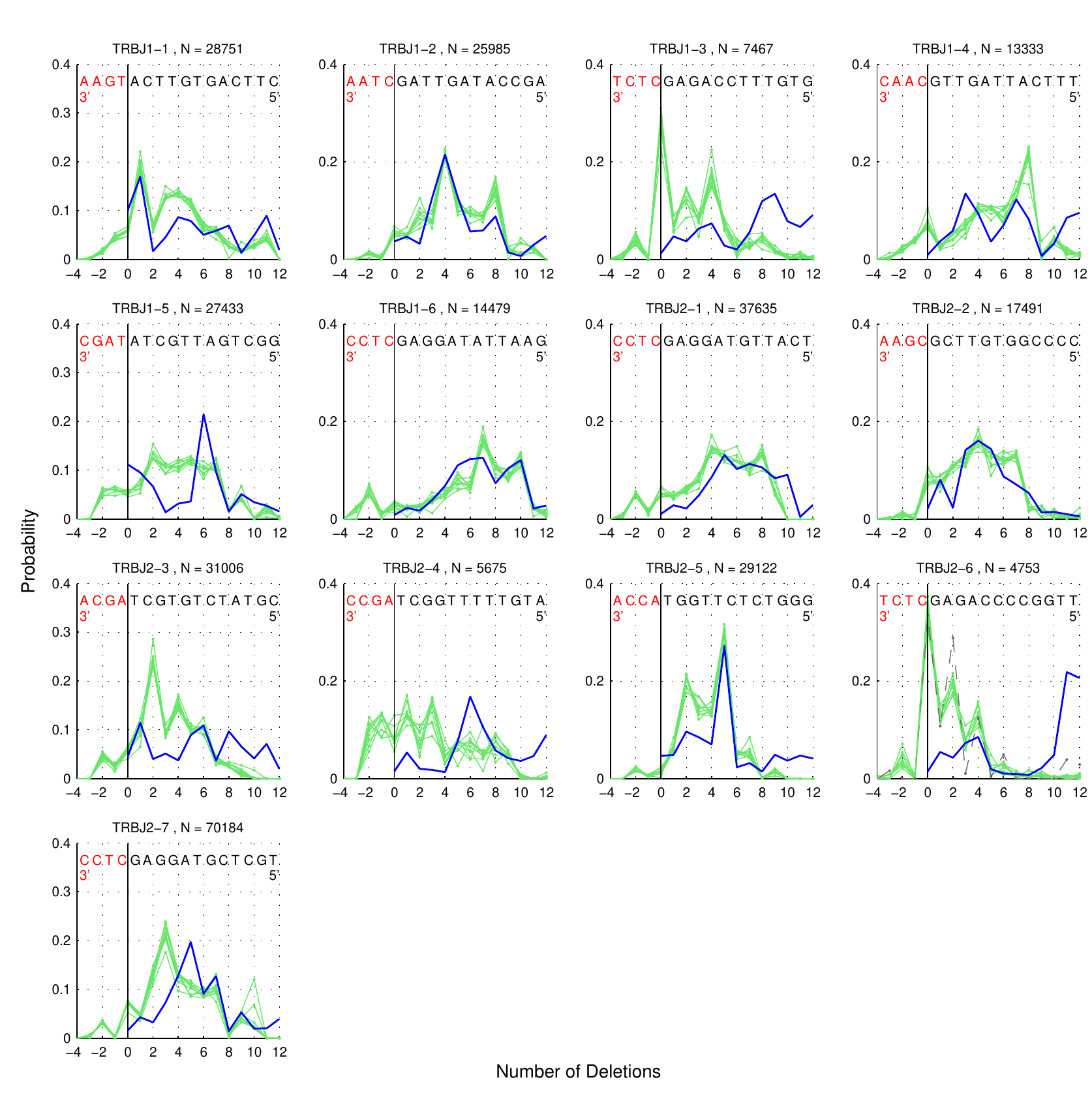}
\end{center}
\caption{Deletion profiles for all the J-genes. The title for each panel lists the gene name and total number of counts, across all the individuals studied, of the particular gene in question. Individuals with fewer than 100 counts for a specific gene are plotted in gray dashed lines. The blue lines show the predictions of the position weight matrix based model fit to the V deletions curves, but evaluated on the J gene sequences.
\label{figgenedeletions4}
}
~
\newpage
~
\newpage

\end{figure*}
\begin{figure*}[htbp!]
\begin{center}
\noindent\includegraphics[width=.9\linewidth]{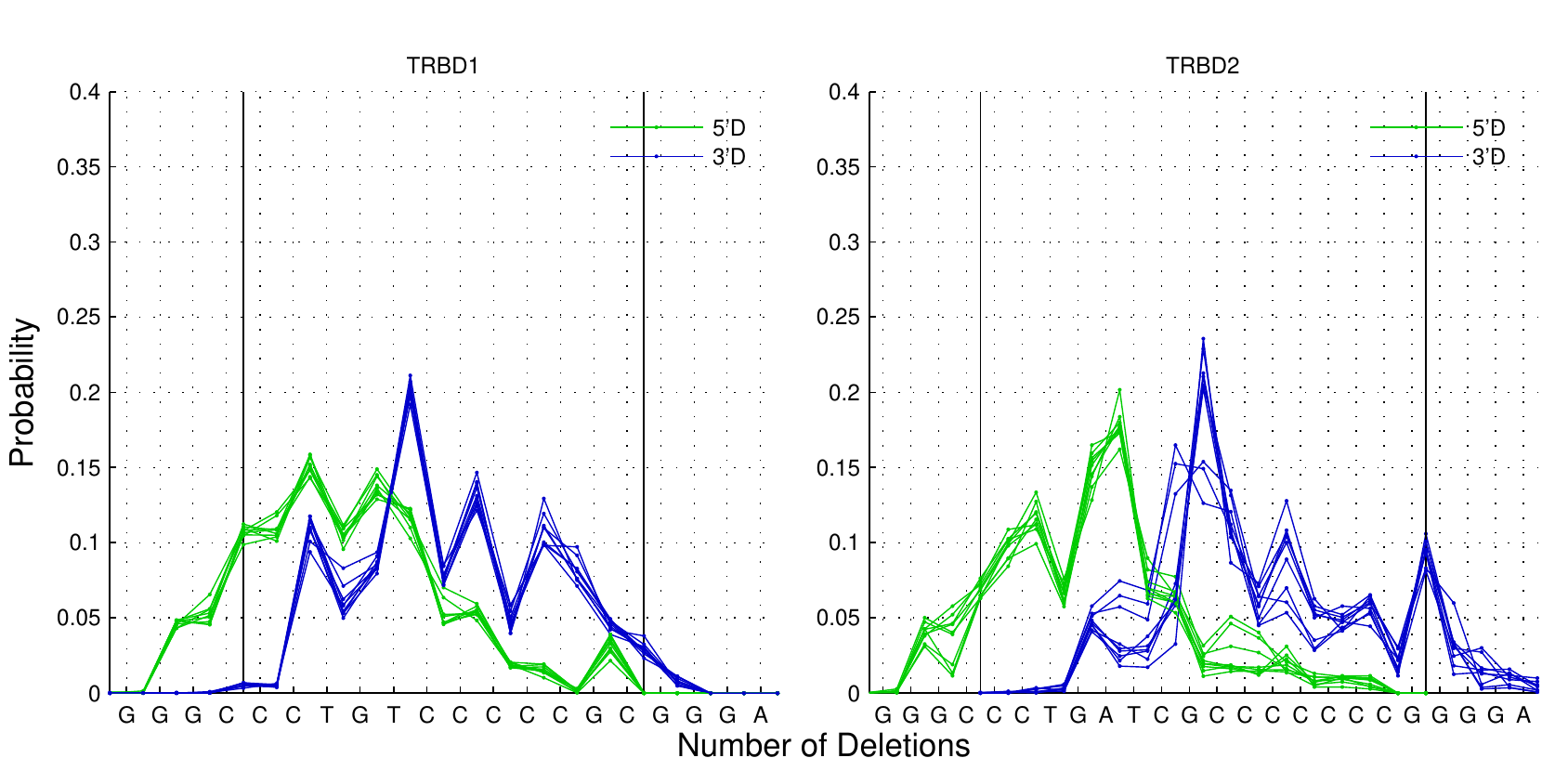}
\end{center}
\caption{Marginal deletion probability distributions for the two D-genes. Deletions at the \fivep end (\threep end) of the D gene are shown in green (blue). The x-axis displays the gene sequence from the \fivep end to the \threep end.
\label{figgenedeletions}
}
\end{figure*}
~
\newpage
~
\newpage
~
\newpage
~
\newpage
~
\newpage

}

%\newpage

\appendix

%\noindent\stardate\\

\section{Sequences of V, D, and J-genes and their alleles}\label{alleles}
Accurate knowledge of the sequences of germ line V-, D-, and J-genes and their allelic variants is essential to minimize errors and bias in our analysis. There are 2 D-genes, 13 J-genes, and 48 V-genes, not counting alleles. There are in addition 19 `pseudo' V-genes on the same germline chromosome: they participate in the recombination process and, though they cannot lead to a functioning receptor, they can appear in the non-productive sequence data sets, provided that a sequencing primer (or an approximate one) is present, which in our case is true for 11 pseudo V-genes. 

We curated a list of known and discovered allelic variants of the V-genes by combining those found in the public IMGT database \cite{Monod:2004im} with variants that we discovered with high confidence during our analysis. Not all the sequence reads listed in IMGT are true variants since many of them are from rearranged DNA with variation at the junctional end. Such `variants' were removed from our list, unless the variation was deeper in the sequence, far from the edited end. 
In addition, we have found three instances of allelic variants in our data that are not listed in IMGT. The discovered variants of genes TRBV7-7 and TRBV10-1 can actually be found by BLAST in the NCBI database of human sequences; the variant of gene TRBV7-2 is not found by BLAST and appears to be completely novel. Undiscovered variants have rather small impact on overall recombination event statistics, but they can cause systematic errors in the inference of gene-specific deletion profiles. 

Complete lists of the genes and alleles used in our analysis are available online \footnote{{\tt \small physics.princeton.edu/\~{}ccallan/TCRPaper/genes}}. For completeness, we also list the primers used by Robins et. al. \cite{Robins:2010hda,Robins:2009da} in acquiring the data we analyze.

\section{CDR3 sequence data files and formats}
The CDR3 sequences used in our analysis come from na\"ive or memory CD4+ T-cells of 9 human individuals, and are further segregated into `in-frame' and `non-productive' sequences. The sequences are 60bp in length for 6 of the subjects, and 101bp in length for the remaining three. The reads of different length differ only in how far the sequencing window goes into the V gene: both types are anchored on the same conserved phenylalanine in the J-gene and have the same read depth into the J-gene. 

Processed sequence data was made available to us by H. Robins. As described in \cite{Robins:2010hda,Robins:2009da} each sequence is read multiple times and the multiple reads are used to estimate the multiplicity of each specific TCR receptor in its respective compartment. In addition, multiple reads are used to correct for sequencing errors by clustering reads that differ at a small number of positions \cite{Robins:2010hda}. In our data files, the effective sequence multiplicity is recorded along with the error-corrected sequence (although we do not use multiplicity in our current analysis). The data files used in our analysis are available online \footnote{{\tt \small physics.princeton.edu/\~{}ccallan/TCRPaper/data}}. The file names in the repository clearly indicate the category to which the included data belongs. 

\section{Overall description of the analysis pipeline and software}\label{overallanalysis}
\panelflowchart
There are two major steps  in the analysis pipeline that leads from a list of CDR3 sequences to a final estimate of the probability distribution $P_{\rm recomb}(E)$ of generative recombination events. The first is an `alignment' step in which, for each read $\sigma$, we create a comprehensive list of recombination \lq scenarios\rq\  $\{E_{\sigma}\}$ that could plausibly have produced that read. A `scenario' is a particular set of values for the event variables (gene identities, VD insertions, etc.) that generates a recombined sequence nearly identical to the read in question (with possibly a small number of mismatches).  The second major step is an iterative procedure (summarized in the flow chart of Fig.\,\ref{figflowchart}) for finding the generative distribution that maximizes the likelihood of the observed data given the functional form of the generative distribution (as expressed in main text Eqn.\,2). 

The algorithms we have developed to execute these two steps are described in greater detail in the following two subsections. Software to implement these procedures was written in Matlab using the Parallel Computing toolbox and run on a Linux cluster.  Compiling key routines into C++ using Matlab Coder greatly improved processing speed, allowing model inference on an individual data set to be completed in about 20 hours running on 8 processors. Our Matlab code, along with summary instructions on how to run it, is available online \footnote{{\tt \small physics.princeton.edu/\~{}ccallan/TCRPaper/scripts}}

\subsection{Initial parsing of sequence reads by alignment}
The first step in our inference procedure is to align each CDR3 read with specific alleles of V, D, and J genes by sequence matching. The goal is to generate a set of plausible recombination events that could produce the read to serve as a starting point for subsequent probabilistic refinement. This preliminary alignment procedure produces, for each read, a finite number of V, D, and J alleles, the maximal length alignments of these alleles to the read, the corresponding minimum nucleotide deletions from the genomic sequences, with possible P-nucleotides identified, and with the unmatched parts of the read identified as VD or DJ insertions. Mismatch information is also stored.

Certain thresholds are imposed on the alignments -- gene alignment lengths must be sufficiently long; gene deletions must not be too large; errors are allowed in the alignments (no gaps), but the number of errors must be small. The alignment score (using an appropriate mismatch penalty) is used to rank order alignments, and a threshold on the score relative to the score of the best alignment is also imposed. Specific values for these various parameters are chosen in the light of computational experience to achieve fast and accurate convergence of the overall model-fitting algorithm. 

The procedure for finding J matches is simplest. The CDR3 reads all begin at the \threep end (sense strand) from a primer in a known position in each J gene. Thus for each candidate J gene, we simply look for exect matches of the end of the sequence read with the portion of the gene just \fivep of the primer. Proceeding in this way, and imposing the various thresholds mentioned, we find an average of 2-3 J alignments per read.

For the V-gene, the position of alignment to the read is not fixed. So for a given V-gene, we align the \fivep end of the read to the m-th base from the \threep end of the V-gene, and note the best-scoring match at this positioning (this time allowing some mismatches, and penalizing them in the score). We step through the values of m and record the best-scoring match over all positionings. Repeating this process for all the V-genes, and imposing the earlier mentioned thresholds, we are left with a limited set of possible V-gene identifications, together with their specific alignments to the read. Proceeding in this way, we find an average of $\sim 15$ V alignments per read.

After identifying the plausible alignments to V- and J- genes, we turn to the problem of identifying D-gene matches. This is a more difficult problem because the D-genes are short, and deletions (occurring on both ends) often leave residual sequences which are hard to identify as a D-gene fragment. We therefore put very loose constraints on the D-gene alignments, relying on the probabilistic refinement to narrow them down. Specifically, we consider the read sequence segment lying between the end of the highest-scoring V-gene and the end of the highest-scoring J-gene, and include 10 nucleotides of flanking sequence on either side, to allow for ambiguous origin of these bases. We identify as a possible D-gene match every maximal non-overlapping alignment to this segment of the three D-gene alleles. These D-gene matches are scored by their length and the top 200 are selected as possible D-gene alignments. 

Alignment files are available online \footnote{{\tt \small physics.princeton.edu/\~{}ccallan/TCRPaper/results/alignments}}: the files are in Matlab format and record the outcome of the above alignment strategy for a subset of our data. Inspection of the alignment data for individual sequences  should provide instructive illustrations of the above-described procedure. The various thresholds and parameters used in the procedure are found in the files as well. The full set of alignment files used in our analysis can be generated using routines provided in our online software repository. 

We note that one could generate a unique assignment of sequence features to a given read by selecting from the alignment ensembles just described the V, D, and J assignments with the highest score (i.e. having the longest effective alignment with the read). We will call the occurrence distribution of gene assignments, insertions, and deletions produced in this way as the \lq deterministic \rq\ estimate of the sequence feature probability distribution. It corresponds to standard practice in the literature for inferring feature statistics from sequence data, and will be used as a benchmark for comparison and contrast with our more accurate probabilistically inferred distribution.

\subsection{The expectation maximization algorithm}
As described in the main text, we wish to find model parameters that maximize the likelihood of the data. We use an iterative Expectation-Maximization algorithm to do this. Given a current guess for the model parameters that describe $P_{\rm recomb}(E)$, we update it by calculating the probability-weighted counts of events over the data set and then using those counts to re-estimate the marginal distributions ($P(V)$, $P(D,J)$, $P({\rm ins}VD)$, and so on) that appear as factors in the general functional form of  $P_{\rm recomb}(E)$ (main text Eqn.\,2). 

As indicated in main text Eqns.\,2-4, the joint likelihood of a recombination event $E$ and sequence $\sigma$ is the product of two factors: the probability of the generative event (given by $P_{\rm recomb}(E)$), and the sum over allele choices $a$ of the probability of those allele choices multiplied by the probability of the number of mismatches between $\sigma$ and the sequence $\sigma_{E}^{a}$ implied by $E$ and $a$. In other words, in addition to the recombination event probability $P_{\rm recomb}(E)$, likelihood involves the sequencing error rate $R$ and the allele probabilities $P(V_a|V)$, etc. We emphasize that we carry out this exercise independently for the data sets derived from different individuals. While we expect (and find) that $P_{\rm recomb}(E)$ is consistent between individuals, we of course expect different individuals to have different allele probabilities.

In the expectation maximization procedure, we start from a prior in which each factor in main text Eqn.\,2 for $P_{\rm recomb}(E)$ is uniform in its variables,  the sequencing error rate $R$ is set to a small value (typically $10^{-4}$), and the allele probabilities are uniform over all the alleles of each gene. Using main text Eqn.\,4, for each CDR3 sequence read $\sigma$, we exhaustively compute the likelihoods of all recombination events $E$ given $\sigma$, starting from maximal alignments for each sequence identified in the initial parsing of the read (previous section), and looping over the other scenarios, involving extra deletions compensated by chance re-insertions of identical nucleotides, that could also \lq explain\rq\ the read. We also loop over the number of true P-nucleotides in the cases where they are present. 

Normalizing these likelihoods yields the relative weights that observing the sequence $\sigma$ assigns to different recombination events $E$, given the current model parameters. Summing these weighted occurrences over all the sequences in the data set gives a new, data-conditioned, estimate of the various factors that enter into the assumed general form of $P_{\rm recomb}(E)$ (as well as a new estimate of the sequencing error probability and allele occurrence frequencies). The formal statement of the update rule is as follows; for each parameter in the model that describes the probability of a specific recombination event feature $X$ (say a particular V-gene choice) we update it to the probability weighted counts over the whole data set of that event. In other words, the $(k+1)$-th iteration of the model parameters are given by
\begin{align}
P^{(k+1)}(X) &= \sum_{\sigma \in \mathcal{D}}  \sum_{E} \delta_{X_{E},X} \, P^{(k)}(E | \sigma) \notag \\
&= \sum_{\sigma \in \mathcal{D}}  \sum_{E} \delta_{X_{E},X} \, \frac{P^{(k)}(E , \sigma) }{L^{(k)}(\sigma)}
\end{align}
where $\delta_{X_{E},X}$ is one if $X$ is true in the recombination event $E$ and zero otherwise. This procedure is used to update all the factors entering into the likelihood calculation and the process is repeated until convergence to a stable end point is achieved. Since all sequences in the data set are looped over in the calculation, we can record \lq on the fly\rq\ the likelihood $L(\sigma)$ (main text Eqn.\,4), the generation probability $P_{\rm gen}(\sigma)$ of that sequence (a conceptually different quantity), as well as the conditional entropy of events $S(E| \sigma)$ for each sequence quantifying the multiplicity of recombination events that could have produced the given CDR3 sequence). The product of $L(\sigma)$ over all sequences is the current overall likelihood of the data set, a measure of convergence of the procedure. The generation probabilities $P_{\rm gen}(\sigma)$ have a direct physical significance, reflecting the probability of generation of the sequence by the molecular machinery.

\panellikelihood

Iterating this process is guaranteed, by general expectation maximization arguments, to maximize the overall likelihood of the data set locally. We have found that rapid and direct convergence to a likelihood maximum is the norm for the data sets we work with (see Fig.\,\ref{figlikelihood}). The models for the probability distribution of generative events inferred in this way from the different data sets are available online \footnote{{\tt \small physics.princeton.edu/\~{}ccallan/TCRPaper/results/models}}. The distribution is also described in a Microsoft Excel file.

\section{Sequencing error rate}
The sequence mismatch rate in our model reflects both uncorrected sequencing error as well as unknown allelic variation. Our model assumes that this mismatch rate $R$ is independent of position along the sequence read. As is well-known, accuracy of the sequencing procedure becomes worse at the end of the sequence read (the \fivep , or V-gene, end of our CDR3 sequence) so, in assaying error rates, we ignore the last 15 nucleotides (at the \fivep end) for the 101 bp reads, where we can afford to do this. Our alignment procedure also disallows mismatches in the J- and D-gene alignment because of the shortness of these segments and the expected low error rate at this end (more accurately, the beginning) of the sequence read. In assessing position dependence of sequence error rates, therefore, we only need concern ourselves with mismatches to V gene assignments. Summing all such mismatches for the three individuals for which we have 101 bp reads, and plotting them against read position, we obtain the results plotted in Fig.\,\ref{figerrorprofile}. We find that $R$ converges in the mean to a value of order $3\times10^{-4}$ per base pair, two orders of magnitude smaller than the raw instrumental sequencing error rate. There are, however, a few sharp peaks at specific positions along the read; since they appear at the same position for different individuals, they presumably reflect some anomaly in the functioning of the sequencing machine. This shortcoming of the error rate model does not greatly influence the results of the inference because the overall error rate is rather low. 
\panelerrorprofile

\section{Gene and pseudogene usage}\label{pseudogene}
\panelgeneusage
In Fig.\,\ref{figgeneusage}, we show the inferred gene usage frequencies. As described in the main text, Fig.\,\ref{figgeneusage}D reveals the mechanistic constraint prohibiting the recombination of the TRBD2 gene with any upstream TRBJ1 gene. We include pseudo V-genes in our analysis. These pseudogenes cannot produce a functional receptor but they can participate in the recombination process and produce a non-productive rearranged CDR3 sequence which can be transmitted into the na\"ive or memory compartments just like any other non-productive rearrangement. The set of V gene sequencing primers used by Robins et. al. \cite{Robins:2010hda,Robins:2009da} either exactly or approximately match 11 pseudogenes. Of these, TRBV23-1, TRBV5-3, TRBV12-2 and TRBV6-7 show significant usage, together accounting for almost 10\% of CDR3 sequence reads.

\section{Memory T-cell non-productive repertoire}\label{memoryprod}

We performed the same analysis on both the naive and memory T-cell repertoires. The non-productive CDR3 sequences in both of these compartments should not be subject to selection, and a comparison of inferences from the two provides a test of this important assumption. Results from the larger na\"ive non-productive compartment (containing an average of 35,000 unique sequences per individual) were reported in the main text. Here we report the results from the smaller memory non-productive compartment (containing an average of 22,000 unique sequences per individual). In Fig.\,\ref{fignaivevsmem}, we compare the naive and memory insertions and deletions distributions. In Fig.\,\ref{figmemoverlap} we show that the occurence of shared sequences between the individual non-productive repertoires is consistent with our generative model for the memory compartments as well. The plots show that the models inferred from the na\"ive and memory T-cells are identical in all respects, in confirmation of the expectation that non-productive sequences are not subject to selection effects.

\panelnaivevsmem

\section{Spurious shared sequences between repertoires}\label{spurious}
Of the 9 individuals, we find three specific pairs of individuals -- (2,3), (2,7) and (5,6) -- who have an unusually large number of sequences in common, in both the naive and memory compartments. While all other pairs of individuals have between 0 and 4 sequences in common, these three pairs have 15 to 90 shared sequences. Additionally, many of these shared sequences occur in both the naive and memory compartments of the individuals. We suspect that these anomalies are the result of inter-sample contamination.

Hence, for our analysis of the distribution of shared sequences between individuals, we discard from consideration the four pairs of individuals (2,3), (2,7), (3,7) and (5,6). This leaves 32 pairs of individuals for our analysis. We also discard three specific additional sequences that occur in the naive and memory compartments of one individual and also in another individual.

\section{Convergent recombination and generation probability}
\panelentropyvspgen
As discussed in the main text, a typical CDR3 sequence can be produced by $\approx 32$ different recombination events, corresponding to an entropy of 5 bits per CDR3 sequence. In Fig.\,\ref{figentropyvspgen}, we show the 2D histogram of the recombination entropy $S(E| \sigma)$  and the generation probability $P_{\rm gen}(\sigma)$. As expected, sequences with higher recombination entropy tend to have higher total generation probability, with a correlation of $0.13$. Note also that while the shared sequences between individuals (red dots) all have high $P_{\rm gen}(\sigma)$, they are widely distributed with respect to the recombination entropy, since only $P_{\rm gen}(\sigma)$ determines the recurrence probability of a sequence.

\section{Generation probabilities of productive sequences}
The probability distribution of recombination events that we infer enables us to calculate the generation probability of any given TCR$\beta$ CDR3 sequence. We calculate $P_{\rm gen}(\sigma)$ for all the sequences in the naive and memory productive repertoires. The distributions of these generation probabilities are shown in Fig.\,\ref{figproductivePgen}. The productive repertoires have systematically higher generation probabilities, implying that sequences that are more likely to be generated are also more likely to pass selection filters and survive in the blood. This is, in part, due to systematically fewer insertions in the productive repertoires, which have exponentially higher generation probabilities.
\panelproductive

\section{Test of analysis on simulated sequences}\label{testsimul}
As noted in the main text, we infer the probability distribution of generative events from nonproductive sequences only. One might worry that using such a non-random subset of all the sequences produced by VDJ recombination could introduce a bias in the inference. We first note that the condition for a rearranged CDR3 sequence to be out of frame involves the sum of six variables that our analysis has shown to be uncorrelated:
\begin{align*}
[&-{\rm del}V + {\rm ins}VD -{\rm del} 5^{\, \prime}D + {\rm length}(D) \\ 
&- {\rm del} 3^{\, \prime}D + {\rm ins}DJ - {\rm del}J] \; {\rm mod} \; 3 > 0.
\end{align*}
Since a large number of uncorrelated variables are involved, it is a priori unlikely that this constraint would significantly influence the evaluation of the distributions that define our generative model. We can test this quantitatively by generating a simulated sequence repertoire from our recombination event distribution, running our inference algorithm on the out-of-frame subset of these sequences, and then comparing the inferred and the "actual" event distributions. The result of this analysis on a simulated repertoire of $10^5$ sequences (two-thirds of which were out-of-frame)  is displayed in Fig.\,\ref{figsimulated}. It is clear that the initial and the inferred generative distributions are identical to each other, confirming that the condition of being out-of-frame does not bias the statistics of recombination events and does not interfere with our ability to correctly infer the probability distribution of these events.

\panelsimulated 

\section{Occurrence of palindromic nucleotides with non-zero deletions}
To show that the occurrence of palindromic nucleotides with non-zero nucleotide deletions from the ends of the genes is consistent with chance insertions, we keep track of the (model probability weighted) joint frequencies of lengths of observed palindromes conditioned on the number of deletions and on gene choice. Keeping track of this detail is necessary because of the strong dependence of deletion probabilities on gene choice. After we obtain our converged model, we calculate the frequencies of chance palindromic nucleotides of different lengths co-occurring with non-zero deletions (taking into account all the structure of $P_{\rm recomb}(E)$, including the nucleotide bias in insertions).  The plot in Fig.\,\ref{figpnuc} shows that the observed frequencies of palindromic nucleotides co-occurring with non-zero deletions are completely consistent with those expected by chance insertions. 

\panelpnuc

\paneldelseqmotif

\panelgenedeletions

\section{Sequence dependence of nucleotide deletion probabilities}\label{delmodel}
Since the sequence at the \threep end of the V gene varies between genes, we fit a simple model to the gene dependent deletions profiles to explain the variation in these distributions. The precise mechanism of the generation of P-nucleotides and their relationship to deletions is unclear. Hence, we take only the probabilities of deletions greater than or equal to two nucleotides and consider the nucleotide sequence context (four bases \threep and two bases \fivep of the deletion position) as a predictor of the deletion probability. We use a function of the form
\begin{equation}
P( n \, \mathrm{deletions} | \sigma \, \& \, n \geq 2) = \frac{\exp \left( \sum_{k=1}^{6} \epsilon(k, \sigma(n - 4 + k)  \right)}{Z(\sigma)},
\end{equation}
\begin{equation}
Z(\sigma) = \sum_{n=2}^{12} \exp \left( \sum_{k=1}^{6} \epsilon(k, \sigma(n-4+k) \right)
\end{equation}
where $\epsilon$ is a $6\times4$ matrix containing the contribution of each possible nucleotide at each of the positions, analogous to a (log) Position Weight Matrix (PWM). We do a least squares fit to determine the elements of $\epsilon$. In Fig.\,\ref{figdelseqmotif}, we show $\epsilon$ fit to the V deletions. There is a strong preference for T and A, especially in the 2 nucleotides just \fivep of the position of deletion. Since there are only 13 J-genes, there is less sequence variation among them that we can utilize.

%\newpage

%\nocite{*}
\bibliographystyle{pnas}

\end{document}